\documentclass[5p, twocolumn]{elsarticle}
\biboptions{sort&compress}

\usepackage{xr-hyper}
\usepackage{hyperref}
\usepackage{cuted}
\usepackage{graphicx,float, amsmath}
\usepackage[dvipsnames]{xcolor}
\usepackage{natbib}
\usepackage{mathrsfs}
\usepackage{subfigure} 
\usepackage[integrals]{wasysym} 
\usepackage{nicefrac,amssymb,amsmath} 
\usepackage{booktabs}
\usepackage{textcomp}
\usepackage{changes}

\newcommand{\qfd} {{\phantom{\dagger}}}

\definecolor{mblau}{HTML}{5E81B5}
\definecolor{morange}{HTML}{E19C24}
\definecolor{mgruen}{HTML}{8FB032}
\definecolor{mrot}{HTML}{EB6235}
\definecolor{mlila}{HTML}{8778B3}
\definecolor{mbraun}{HTML}{C56E1A}
\definecolor{mhellblau}{HTML}{5D9EC7}
\definecolor{maubergine}{HTML}{A5609D}
\definecolor{mkhaki}{HTML}{929600}
\definecolor{mblassrot}{HTML}{E95536}
\definecolor{mittelblau}{HTML}{6685D9}
\definecolor{morange2}{HTML}{F89F13}
\definecolor{mgrau}{HTML}{808080}

\journal{Physica E}

\begin{document}

\begin{frontmatter}

\title{Dynamic correlations in the highly dilute 2D electron liquid: loss function, critical wave vector and analytic plasmon dispersion}

\author{J\"urgen T. Drachta}
\author{Dominik Kreil\corref{corA}}
\author{Raphael Hobbiger, and Helga M. B\"ohm}
\address{Institut f\"ur Theoretische Physik, Johannes Kepler University, 4040 Linz, Austria}
\cortext[corA]{Corresponding author}




\begin{abstract}

Correlations, highly important in low--dimensional systems, are known to decrease the plasmon dispersion of
two?dimensional electron liquids. Here we calculate the plasmon properties, applying the ?Dynamic Many-Body
Theory?, accounting for correlated two-particle--two-hole fluctuations. These dynamic correlations are found to
significantly lower the plasmon's energy. For the data obtained numerically, we provide an analytic expression
that is valid across a wide range both of densities and of wave vectors. Finally, we demonstrate how this can be
invoked in determining the actual electron densities from measurements on an AlGaAs quantum well.
\end{abstract}


\end{frontmatter}


\section{Introduction}

The study of plasmon excitations in  electron systems traces back 80 years, to Wood's observation~\cite{wood1933remarkable} of a characteristic reflectivity drop in alkali metals. 
Plasmons excited by electrons impinging on metals were found 15 years later~\cite{lang1948geschwindigkeitsverluste, ruthemann1948diskrete}, and soon after explained by Bohm and Pines~\cite{bohm1953collective} 
with their mean field or `random phase' approximation (RPA). 
When manufacturing of high-quality semiconductor- and metal interfaces became possible, the two--dimensional electron liquid (2DEL) provoked attention~\cite{allen1977observation}.
Electrons confined to a He surface remain another quintessential 2DEL~\cite{AGHK16:spectroscopy}.

RPA calculations of plasmons in single- and double-layer graphene were performed in Refs.\ \cite{Sarma13:intrinsic,TaHo14:effects_of_temp,VanTuan13:plasmon_modes} (with references to earlier work), which all included temperature effects.
For the novel 3D Dirac liquids in semimetals, such as Na$_3$Bi
, the RPA plasmon was studied in Ref.\ \cite{HofD15:Plasmonsignature}; massive Dirac particles were treated in Ref.\ \cite{ThSA17:Dynamic}.
For recent work on 1D plasmons we refer to \cite{Grosu:2008}.

Angle-resolved photoemission spectra, containing periodic crystal as well as
many-electron effects, also clearly show a plasmon's fingerprint, essentially
probing the single--particle propagator's `spectral function' \cite{giuliani2005quantum}.
Pertinent work for 2DELs is found in \cite{Losic14:Spectral_2D, VFSL16:Dispersion, PABB08:Plasmons}.

Premium data directly on 2D plasmons were obtained by Nagao et al.~\cite{nagaoplasmon, RuNP08:experimental}, who studied  the sheet plasmon in Ag surface state bands on Si and in DySi$_2$ monolayers on Si both with high resolution electron energy loss spectroscopy (HREELS), and by Hirjibehedin et al.\ \cite{hirjibehedin2002evidence,EPDH00:collective} for AlGaAs quantum wells (QWs) using inelastic light scattering.
The former group measured a 2DEL of moderate areal density
\(n\!=1/\pi(a_{\scriptscriptstyle\rm B}^*r_\mathrm{s})^2\) with
\(r_{\mathrm{s}}\!\lesssim\!2\) (\(a_{\scriptscriptstyle\rm B}^*\) is the
effective Bohr radius),
while the QW-2DELs were rather dilute with
\(n\!\approx2\!\times\!10^{13}\ldots8\!\times\!10^{8}\mathrm{cm}^{-2}\) (\(r_\mathrm{s}\!\approx 10\ldots20\)).

When the ratio kinetic to potential energy decreases, correlations get increasingly important.
They play a significant role in the above low density QWs (in contrast to dense oxide--interface electron gases \cite{HaoDiebold2015Coexistence, FarA17:plasmons}, which are well described by the RPA).
The dilute electron liquids require correcting the RPA's local field for the exchange--correlation hole,
which changes dynamically.
For perturbations with wavelengths as low as the interparticle distance, this is crucial.
The \textit{Dynamic Many Body Theory} of Krotscheck et al.\ \cite{BHKP10:dynamic,CaKL15:dynamic, Halinen:2003} has proven excellent in this regime.
The fermion version includes dynamically coupled 2-particle\,--\,2-hole (2p2h) excitations.
We here use it to study the 2DEL, focusing on the plasmon.

Beside the correlations, the layer width \(L\) acts to decrease the plasmon energy (as the Coulomb interaction is better screened than in the strictly 2DEL).
Higher temperatures \(T\), attenuating the interaction-to-kinetic-ratio, similarly diminish correlation effects.
In the plasmon dispersion of typical semiconductor QWs all these influences can mutually cancel \cite{HwaS01:plasmon},
resulting in a `classical' \(\sqrt{q}-\)plasmon dispersion (\(q\) denotes the wave vector,
\(\omega\) the frequency).

Our aim here is a state-of-the-art calculation of the correlation contribution to the plasmon properties.
We also present a genuine two-dimensional fit of the numerical results in the \((q,r_{\scriptscriptstyle\mathrm S})-\)plane, for comparison with other works and applications.
In order to clearly bring out where correlation effects can become important, \(T\) and \(L\) are mostly kept zero.

Our work is organized as follows: 
In Sec.\ \ref{sec: GRPA} we investigate the plasmon dispersion including static electron correlations,  
using two models both based on the most accurate available simulation data \cite{davoudi01:analytical, gori2004pair}.
The dynamic 2p2h theory and its underlying physics are briefly introduced in Sec.\ \ref{sec: Dynamic Many Body Theory},
our numerical results for the 2D plasmon together with the analytic fit being presented in Sec.\ \ref{sec: Results of the 2p2h Theory}.
In Sec.\ \ref{sec: Plasmon dispersion in semiconductor QWs} we first adapt the expression to realistic QWs and then apply our approach to determine the electron density of experimental samples, followed by our conclusions in Sec.\ \ref{sec: Conclusions}.

\section{Theories of a G(eneral)RPA type \label{sec: GRPA}}

The density response of an electron gas to an external potential
\(V_\mathrm{\!ext}(q,\omega)\) defines its linear response function \(\chi\,\), 
\begin{subequations}
 \begin{equation}
  \delta \rho(q, \omega) = \chi(q, \omega)\, V_\mathrm{ext}(q, \omega)
  \;,
 \end{equation}
or, equivalently, the dielectric function \(\epsilon\,\), via
 \begin{equation}
  \epsilon^{-1}(q, \omega) = 1 + v(q)\,\chi(q, \omega) 
  \;.
 \end{equation}
 \label{eq: def chi eps}
\end{subequations}
Denoting the response of non--interacting fermions as \(\chi^0(q,\omega)\)
and the Coulomb interaction as \(v(q)\),
the exact response in Eq.\,\eqref{eq: def chi eps} leads to \(G(q,\omega)\) via
\begin{subequations}
 \begin{equation}
  \chi =\, 
    \frac{\chi^0}{1-v\,(1\!-\!G)\,\chi^0} \;,
  \label{eq: def chi G}
 \end{equation}
 \begin{equation}
  \epsilon=\, 1-\, \frac{v\chi^0}{1+G\,v\chi^0} \;.
  \label{eq: def eps G}
 \end{equation}
 Comparison with the Clausius-Mossotti form \(\epsilon\!= 1\!+\widetilde\alpha/(1\!-\!\frac13\widetilde\alpha)\)
 in solids showing a molecular polarizability \(\epsilon_0\widetilde\alpha/n\), explains the name `local field correction' (LFC) for \(G\) \cite{polinitosiMB06}.
 If the interaction has no Fourier transform, (e.g.\ dipoles or hard-core particles), it is preferable to define a dynamic effective 
interaction \(V_{\!\epsilon}(q,\omega)\),
 \begin{equation}
  \chi =\, 
    \frac{\chi^0}{1-V_{\!\epsilon}\,\chi^0} \;,
    \label{eq: def Vqweps}
 \end{equation}
 \label{eq: GRPA}
\end{subequations}

By choosing \(G(q,\omega) =\!0\), one recovers the bare RPA.  It shows two main features, the particle--hole band (PHB), and an undamped plasmon:
\begin{subequations}
 \label{eq: RPA}
 \begin{equation}
  \chi^{\scriptscriptstyle\mathrm{RPA}}(q,\omega) = \>
   \frac{\chi^0(q, \omega)}{1-v(q)\,\chi^0(q, \omega)}  
 \label{eq: chi_RPA}
 \end{equation}
 \begin{equation}
  \epsilon^{\scriptscriptstyle\mathrm{RPA}}(q,\omega) = \>
   1-v(q)\,\chi^0(q, \omega)  
 \label{eq: eps_RPA}
 \end{equation}
\end{subequations}
For high densities this describes plasmons well, however, it massively overestimates
their energy for dilute systems.

The use of a static \(V_{\!\epsilon}(q)= v(q)\,(1\!-\!G(q))\), termed here GRPA, allows to go beyond the bare RPA, while still retaining its formal simplicity. 
For static perturbations \(G(q)\) coincides with \(G(q,0)\). 
Davoudi et al.\ \cite{davoudi01:analytical} derived its analytical expression in the 2DEL up to \(r_\mathrm{s}\!\le\!10\), based on quantum Monte Carlo (QMC) data for \(\chi(q,0)\) \cite{MoCS92:Static_response} and accounting for the exact limits.
The relation to the Fourier transform of the exchange--correlation
kernel \(f_{\mathrm{xc}}\) in density functional theory is given by
\begin{equation}
 f_{\mathrm{xc}}(q) =\> \left\{\!\begin{array}{lll}
    -v(q)\>G(q,0) \vspace{0.1cm}\\
    -v(q)+V_{\!\epsilon}(q,0) \end{array}\right.
 \label{eq: fxc}
\end{equation}

A different choice of \(G(q)\) is motivated by scattering experiments.  
The fluctuation--dissipation theorem relates the loss--function, \(-\mathrm{Im}\,\chi(q,\omega) \propto \mathrm{Im}\,\epsilon(q,\omega)\), to the van Hove dynamic structure factor \(S(q,\omega)\); 
this, in turn, determines the double differential scattering cross section:
\begin{equation}
 -\hbar\,N\, \mathrm{Im}\chi(q,\omega)
 \>=\>  \pi \, S(q, \omega) \>\propto\> 
 \frac{d^2 \sigma}{d\Omega\,d\hbar\omega} \;,
 \label{eq: fluct-diss}
\end{equation}
(\(\Omega\) is the solid angle, the prefactors depend on the type of measurement). 
The energy--integrated spectrum then yields the static structure factor,
\begin{equation}
 S(q) =\> -\frac{1}{\pi}\!\int\limits_0^\infty\!d(\hbar\omega)\>
 \mathrm{Im}\, \chi(q,\omega)
 \label{eq: 0.moment}
\end{equation}
%
%
(0th moment sum rule). The (static) `particle--hole potential' \cite{KroTriesteBook}
is defined to fulfill this relation, 
\begin{equation}
 - \frac{1}{\pi}  \int\limits_0^\infty\!d(\hbar\omega)\>
   \mathrm{Im}\, \frac{\chi^0(q,\omega)}{1-V_{\!_{\mathrm{ph}\!}}(q)\,\chi^0(q,\omega)} 
    \>=\> S(q) \;;
 \label{eq: chi_GRPA Vph}
\end{equation}
the corresponding LFC is obtained via \(V_{\!{\scriptscriptstyle\mathrm{ph}\!}} \equiv
v\,(1\!-\!G_{\!{\scriptscriptstyle\mathrm{ph}\!}}) \).
For many purposes \(V_{\!{\scriptscriptstyle\mathrm{ph}\!}}\) is well approximated by
\begin{subequations}\label{eq: Vph0 and X}\begin{align}
 V^0_{\!{\scriptscriptstyle\mathrm{ph}\!}}\; &\!=\> \phantom{+}
  \frac{\hbar^2q^2}{4m}\Big[\frac1{S(q)^2}\!-\!\frac1{S^0(q)^2} \Big]
  \label{eq: Vph0}
 \vspace{0.2cm}\\ &\!\equiv\>
  -\frac{\hbar^2q^2}{4m}\,\Big[\frac1{S(q)}\!+\!\frac1{S^0(q)}\Big] \,X(q)
  \;, 
  \label{eq: Xdef}
\end{align}
\end{subequations}%
where \(S^0(q)\) denotes non-interacting fermions and \(X(q)\) the 'direct correlation function'.

The Fourier transform of \(S(q)\) gives the pair distribution function, where, again, fits of state-of-the-art QMC data are available \cite{gori2004pair, kreil2015excitations}.
Clearly, the such defined \(G_{\scriptscriptstyle\!\mathrm{ph}\!}(q\!\to\!\infty)\) cannot diverge, as required for \(G(q\!\to\!\infty,0)\), 
and appears more apt for usage with a Niklasson \(\chi^0(q,\omega)\) \cite{SeMC96:thelocal, Nikl74:dielectric}.

A large variety of other static \(G(q)\) exists \cite{giuliani2005quantum}; 
for recent work on finite-width 2DELs c.f.\ \cite{Bhukal:2015} 
and 
\cite{Aharonyan:2011}.
We here stick to \(G(q,0)\) and \(G_{\scriptscriptstyle\!\mathrm{ph}\!}(q)\) as these LFCs are based on high-quality simulation data. 

For long wavelengths the exact and the RPA static structure factor of a 2DEL
obey \cite{Iwam84:sum_rules} (all \(c,d\) are constant)
\begin{subequations}
 \begin{equation}
  S(q) \>\xrightarrow{q\to0}\> c_{_\mathrm{pl}}\,q^{3/2} +\, c_{_\mathrm{1ph}} \,q^{3} +\, c_{_\mathrm{2p2h}} \,q^{4} \;,
  \label{eq: Sqto0 tru}
 \end{equation}
with \(c_{_\mathrm{pl}}\!= \sqrt{a^*_{\scriptscriptstyle\mathrm{B}}/8\pi n} \,\), and
 \begin{equation}
  S^{\scriptscriptstyle\mathrm{RPA}}(q) \>\xrightarrow{q\to0}\> 
    c_{_\mathrm{pl}}\,q^{3/2}\big(1\!+ d^{\scriptscriptstyle\mathrm{RPA}}_{_\mathrm{pl}}\,q\big) +\,
    c^{\scriptscriptstyle\mathrm{RPA}}_\mathrm{1ph} \,q^{3}  \;.
  \label{eq: Sqto0 RPA}
 \end{equation}
 \label{eq: Sqto0}
\end{subequations}
The leading term arises from the classical \(\sqrt{q}\,\)-plasmon,
\begin{equation}
 S(q) \>\xrightarrow{q\to0}\> \frac{\hbar q^2}{2m\, \omega_{0\!}(q)}\,; \quad
 \omega_0^2 \equiv 
   \frac{2\pi e^2n }{m\,\varepsilon_{\scriptscriptstyle\mathrm{b}}}\,q
 \label{eq: wp0_def}
\end{equation}
(\(\varepsilon_{\scriptscriptstyle\mathrm{b}}\) is the background dielectric constant and \(m\) the effective mass).
With decreasing \(r_{\scriptscriptstyle\mathrm{S}}\) the exact \(S(q)\) must
approach that of the RPA, containing \(q^{5/2}\).
According to \eqref{eq: Sqto0 tru} arbitrarily many particle--hole pairs yield higher order contributions only.
We therefore expect such a term 
to arise from the plasmon also in dilute systems.

The poles of the response function,  Eq.\,\eqref{eq: def Vqweps}, determine the plasmon's dispersion, \(\omega_\mathrm{pl}(q)\).  
All static LFCs yield a mode outside the PHB with
\begin{align}
 \omega_\mathrm{pl} &=\; \omega_{0\!}(q)\,
   \Big(1+\frac{E_{\scriptscriptstyle\mathrm{F}}}{V_{\!\epsilon}(q)}\Big)
   \Big(\frac{V^2_{\!\epsilon}(q)/v(q)}{V_{\!\epsilon}(q)+E_{\scriptscriptstyle\mathrm{F}}/2} +
        \frac{\hbar^2q^2}{4m\,v(q)}\Big)^{\!1/2}
   \label{eq: wpl_GRPA}
   \\
   &\approx\; \omega_{0\!}(q)\,
    \Big(1+\frac{3E_{\scriptscriptstyle\mathrm{F}}}{4V_{\!\epsilon}(q)}\Big)
    \Big(\frac{V_{\!\epsilon}(q)}{v(q)} \Big)^{\!1/2} \;.
   \label{eq: wpl_GRPA qto0}
\end{align}
The compressibility sum rule for \(\epsilon(q\!\to\!0,0)\) \cite{giuliani2005quantum} requires that for any static LFC
\begin{equation}
 V_{\!\epsilon}(q\!\to\!0) \>=\> v(q) + \frac1{n\kappa} - \frac1{n\kappa^0} \;,
 \label{eq: kappa_SR V}
\end{equation}
where \(\kappa^{0}\) is the compressibility of the (free) system.  
This implies the long wavelength plasmon dispersion
\begin{equation}
 \omega^{\!^{\mathrm{GRPA}}\!\!}_\mathrm{pl}(q\!\to\!0) \>=\> 
 \omega_{0\!}(q)\,\Big(1+
   \frac{qa_{_{\mathrm{B}}}^*}{8}\big(1\!+\frac{2\kappa^0}{\kappa}\big)\Big) \;.
 \label{eq: kappa_SR omp}
\end{equation}
Due to finite size effects, QMC calculations cannot provide data for \(q\!\to\!0\).
The plasmon dispersion being highly sensitive to small changes in \(V_{\!\mathrm{ph}}\),
we therefore corrected the fit of Ref.\ \cite{gori2004pair} to ensure Eq.\,\eqref{eq: kappa_SR omp}.

\begin{figure}[H]
 \centering
 \includegraphics[width=0.23\textwidth]{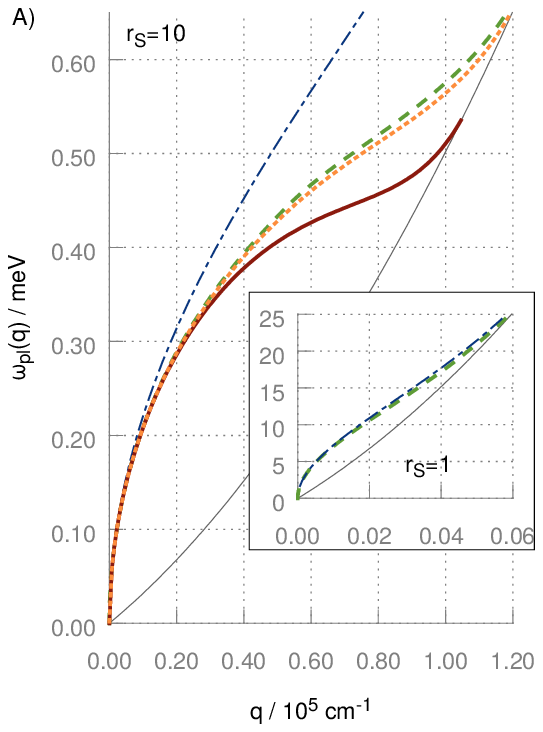}
 \hfill
 \includegraphics[width=0.21\textwidth]{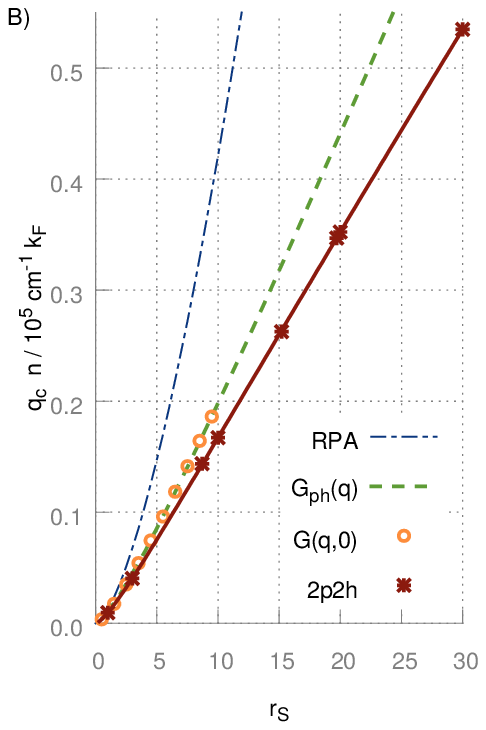}
 \caption{\label{fig: comparedisp and qc}
  Plasmon dispersion (left) and critical wave vector (right) in different theories: bare RPA (dash-dotted blue lines), 
   GRPA with $G_{\!\scriptscriptstyle\mathrm{ph}}(q)$ (dashed green lines) and
with $G(q,0)$ \cite{davoudi01:analytical} (orange short-dashed line (left) and, for a better distinction, orange circles (right)).
   The dark red results are from the dynamic pair theory of Sec.\,\protect{\ref{sec: Dynamic Many Body Theory}}.
}
\end{figure} 

In Fig.\,\ref{fig: comparedisp and qc} (left part) we compare \(\omega_\mathrm{pl}(q)\) obtained with \(G(q,0)\) and with \(G_{\!\scriptscriptstyle\mathrm{ph}}(q)\) for \(r_{\scriptscriptstyle\mathrm S}\!=\!10\). 
The agreement is amazing. The former approach uses \(\omega\!=\!0\) data to describe an \(\omega\!\gg qv_{\scriptscriptstyle\mathrm F}\) mode, while the latter is based on the \(\omega-\)integrated excitations to describe a single point \(\omega_\mathrm{pl}(q)\). The inset shows the plasmon dispersion with RPA as well as \(G_{\!\scriptscriptstyle\mathrm{ph}}(q)\) for \(r_{\scriptscriptstyle\mathrm S}\!=\!1\). In this density regime, theories beyond RPA does not show much improvement. 
The critical wave vectors \(q_c\) for Landau damping, again, almost coincide for all densities (where \(G(q,0)\) is available).
In the \(G_{\!\scriptscriptstyle\mathrm{ph}}(q)\) approach \(q_c\) measured in \(k_{\scriptscriptstyle\mathrm F} = \sqrt{2\,\pi\,n}\,\) flattens around \(2k_{\scriptscriptstyle\mathrm F}\) for large \(r_{\scriptscriptstyle\mathrm S}\) (equivalent to a linear slope the units chosen).

Certainly, local field corrections massively lower the plasmon dispersion from its bare RPA value (blue lines in Fig.\,\ref{fig: comparedisp and qc}). 
Finite \(T\) and \(L\) effects, acting in different directions \cite{HwaS01:plasmon}, cannot be expected to cancel the combined many-body correlations for all \((r_{\scriptscriptstyle\mathrm S},q)-\)combinations.

The dispersion and thus \(q_c\) are further significantly lowered by dynamic correlations (dark red lines in the figure).
We therefore discuss the underlying theory next.

\section{Dynamic Many Body Theory \label{sec: Dynamic Many Body Theory}}

All (static G)RPA approaches, Eqs.\,\eqref{eq: RPA}--\eqref{eq: GRPA}, give no plasmon broadening outside the PHB.
Scattering by impurities and phonons is beyond the jellium model; the lifetime \(\tau\!\equiv\!\gamma^{-1}\) is often treated via replacing \(\chi^0(q,\omega) \to \chi^{0\gamma}(q,\,\omega,\,i\gamma)\)
(Lindhard-Mermin function \cite{Merm70:Lindhard}). A significant group of dynamic LFCs are of the so--called ``quantum STLS'' type \cite{giuliani2005quantum}.
In 3D these approaches describe the plasmon poorly \cite{HolR87:dynamic_qSTLS3D}, yielding \(\mathrm{Im}\,\epsilon \propto -q^5/\omega^7\) near \(\omega_{\mathrm{pl}}\) instead of the exact \(+q^2/\omega^{11/2}\).
We therefore refrain from discussing these theories further.  
(We are not aware of an analogous analytic 2D investigation, for a thorough numerical study, including finite width and finite \(T\) effects, see \cite{yurtsever2003dynamic}. These authors also study the dilute 2DEL in coupled bilayers \cite{TasT10:plasmonic}.).

Intrinsic damping via multi--pair excitations requires a \(q-\)dependent lifetime and  intricate response functions.
A cornerstone, treating dynamic correlations, was presented by Neilson et al.\,\cite{NSSS91}.  Their density response function has the formal structure
\begin{subequations}
 \label{eq: NSSS}
 \begin{equation}
   \chi^{\scriptscriptstyle\mathrm{NSSS}} \;=\; \frac{\chi^{0\gamma}}
    {1- \big[V_{\!_\mathrm{ph}} - \frac{m\omega}{q^2}\,
                                  (\gamma\!-\!\gamma^\mathrm{s})\big]\,\chi^{0\gamma}}
  \;, \label{eq: NSSS_chi}
 \end{equation}
 \begin{equation}
   \epsilon^{\scriptscriptstyle\mathrm{NSSS}} =\, 1\!- \frac{v\,\chi^{0\gamma}}
    {1+ \big[G_{\!_\mathrm{ph}}\!
      - \frac{m\omega}{\omega^2_0(q)}
                                    (\gamma\!-\!\gamma^\mathrm{s})\big]\,v\,\chi^{0\gamma}}
  \;, \label{eq: NSSS_eps}
 \end{equation}
\end{subequations}
where \(\gamma(q,\omega)\) is a mode-mode coupling memory function and
\(\chi^{0\gamma}\) is the Lindhard-Mermin function with the constant \(\gamma\) 
replaced by the ,,self-motion'' function \(\gamma^\mathrm{s}(q,\omega)\).

The Dynamic Many Body Theory \cite{BHKP10:dynamic} accounts for correlated 2-particle\,--\,2-hole (2p2h) excitations.
Its strength lies in incorporating the best available static properties while determining the dynamic correlations via optimization.  
The derivation is sketched in \ref{app: 2-Pair Fluctuations}\, and yields
\begin{subequations}
 \label{eq: 2p2h}
 \begin{equation}
  \chi^{\scriptscriptstyle\mathrm{2p2h}} \;=\; \frac{\Pi_{\mathrm s}}
   {1- \big[V_{\!_\mathrm{ph}} + V_{\!_\mathrm{2p2h}} \big]\,\Pi_{\mathrm s} }
 \;, \label{eq: 2p2h_chi} \end{equation}
 \begin{equation}
  V_{\!_\mathrm{2p2h}} \>=\>
    \textstyle\frac14 \big(\frac1{S^2}\!-\!\frac1{{S^0}^2}\big)\,\displaystyle
    \big(W^{\scriptscriptstyle+}_{\mathrm{s}}\!+\!W^{\scriptscriptstyle-}_{\mathrm{s}}\big)
  \;, \label{eq: 2p2h_V} \end{equation}
 and
 \begin{equation}
  \epsilon^{\scriptscriptstyle\mathrm{2p2h}} =\, 1\!- \frac{v\,\Pi_{\mathrm s}}
   {1+ \big[G_{\!_\mathrm{ph}\!} + G_{\!_\mathrm{2p2h\!}}\big]
       \,v\,\Pi_{\mathrm s} }
  \;, \label{eq: 2p2h_eps} \end{equation}
 \begin{equation}  \phantom{\Big|^|}\hspace{0.8cm}
  G_{\!_\mathrm{2p2h\!}} =\> \big(G^0_{\!_\mathrm{ph}\!}\!-\!1\big)
   \textstyle\frac{m}{\hbar^2q^2}\displaystyle
   \big(W^{\scriptscriptstyle+}_{\mathrm{s}}\!+\!W^{\scriptscriptstyle-}_{\mathrm{s}}\big)
  \;. \label{eq: 2p2h_LFC} \phantom{\bigg|_|}\end{equation}
\label{eq: 2p2h}
\end{subequations}
The `single--particle\footnote{Note that for interacting systems this distinction is ambiguous.} polarizability'
\(\>\Pi_{\mathrm s}\!=\Pi_{\mathrm s}^{\scriptscriptstyle+}\!+\!\Pi_{\mathrm s}^{\scriptscriptstyle-}\>\) with
\begin{subequations}
 \begin{equation}
  \Pi_{\mathrm s}^{\scriptscriptstyle\pm} \>=\>
   \frac{\chi^{0\scriptscriptstyle\pm} }
        {1- W^{\scriptscriptstyle\pm}_{\mathrm{s}} \chi^{0\scriptscriptstyle\pm}}
  \;, \label{eq: Pis} \end{equation}
 \begin{equation} \vspace{-0.0cm}
   W^{\scriptscriptstyle\pm}_{\mathrm{s}} =\>
    \textstyle\frac12\big(1\!+\!\frac{S}{S^0}\big)\displaystyle W^{\scriptscriptstyle\pm}  +
    \textstyle\frac12\big(1\!-\!\frac{S}{S^0}\big)\displaystyle W^{\scriptscriptstyle\mp}  
  \phantom{\Big|_|}\hspace*{-0.5cm} \label{eq: Wspm} \end{equation}
\end{subequations}
builds on the absorption and emission parts of \(\chi^0\), 
\begin{equation}
 \chi^{0\scriptscriptstyle\pm}(q,\omega) =\> 
 \frac1{N}\!\sum_h \frac{n_h (1\! -\! n_{h+q})}
      {\hbar\,\omega \pm (\varepsilon_h \!-\! \varepsilon_{h+q}) + i0^+} \;,
 \label{eq: chi0pm}
\end{equation}
(\(h\) includes the spin index) and the dynamic interactions,
\begin{equation}
  W^{\scriptscriptstyle\pm\!}(q,\omega) \>=\> \frac1{2N}\!\!\sum\limits_{\mathbf{q}', \mathbf{q}''} 
   \delta_{{\bf q},\,{\bf q}'+{\bf q}''}\, \big|\bar K_{\mathbf{q},\mathbf{q'},\mathbf{q''}} \big|^2\,
   \mu^{\scriptscriptstyle\pm\!}(q',q'',\omega) 
  \;. \label{eq: Wpmdef}
\end{equation}
We will no longer spell out the momentum conservation (but use \({\bf q}''\) as abbreviation for \({\bf q}\!+\!{\bf q'}\)).
The pair propa\-gator has, again, a mode-mode coupling structure: Its absorption part \(\mu^{\scriptscriptstyle+}\)  (an analogous form holds for \(\mu^{\scriptscriptstyle-}\)) is
\begin{equation}
 \mu^{\scriptscriptstyle+\!}(q',q'',\omega) \>=\>
 -\!\int\limits_{-\infty}^\infty\!\! \frac{\mathrm{d}\hbar\omega'}{2\pi i}\; 
   \varkappa^{\scriptscriptstyle+\!}(q', \omega')\,
   \varkappa^{\scriptscriptstyle+\!}(q'',\omega\!-\!\omega')
 \;, \label{eq: mudef}
\end{equation}
with
\begin{equation}
 \varkappa^{\scriptscriptstyle+} \>=\> 
 \frac{\chi^{\scriptscriptstyle0+}\, S^2 }{ {S^0}^2 + \hbar\omega\,S^0S X\,\chi^{\scriptscriptstyle0+}}
 \;. \label{eq: kappa+} 
\end{equation}
The function
\(\varkappa^{\scriptscriptstyle+}\!+\!\varkappa^{\scriptscriptstyle-}\) closely resembles \(\chi^{\scriptscriptstyle\mathrm{GRPA}}\).
In particular, their \(\omega^0-\) and \(\omega^1-\)moments agree and their collective modes are also well matched \cite{BHKP10:dynamic}. 
In the plasmon-pole approximation (PPA, also termed `collective approximation')
they are identical, see \ref{app: Collective Approximation}, Eq.\,\eqref{eq: kapppmCA}.

The non--nodal (quantum Ornstein-Zernike) function \(X(q)\), Eq.\,\eqref{eq:
Xdef}, is the main ingredient for the three--body vertex in the dynamic interactions \eqref{eq: Wpmdef}, together with equilibrium triplet correlations \(\bar u^{(3)}\) \cite{BHKP10:dynamic},
\begin{equation}\begin{array}{lllll}\displaystyle\!\!\!\!
 \bar K_{\mathbf{q},\mathbf{q'},\mathbf{q''}} &\displaystyle\!\!\!=\> 
  \mbox{\small\(\displaystyle\frac{\hbar^2}{2m}\)}\,\Big[\,
  \mathbf{q}\!\cdot\!\mathbf{q}' \,X(q')  \,+\,
  \mathbf{q}\!\cdot\!\mathbf{q}''\,X(q'') 
  \vspace{0.2cm}\\ &\displaystyle\hfill -\,\; q^2\;
       \bar u^{(3)}_{\mathbf{q},\mathbf{q'},\mathbf{q''}} \,\Big]\>.\!\!\!\!
  \end{array}
  \label{eq: vertexK}
\end{equation}

\section{Results of the 2p2h Theory \label{sec: Results of the 2p2h Theory}}

For very short-lived plasmons caution is in order \cite{HDKB17_Phenomenplbroadening} whether they are defined as the real part of the complex zero of \(\epsilon(q,z)\) with \(z\!\equiv\omega\!+\!i \Gamma/2\), or as the maximum of the loss function,
\begin{subequations}
 \begin{align} \phantom{\big|_|}
   \epsilon(q,\,z_\mathrm{pl}) &= 0 \;,
  \label{eq: plasmon cmplx0}
  \\
   -\mathrm{Im}\,\epsilon^{-1}(q, \omega_\mathrm{pl}) &\rightarrow \mathrm{max} \;.
  \label{eq: plasmon max}
\end{align}
\end{subequations}
For comparing calculated plasmon positions \(\omega_\mathrm{pl}(q)\) with HREELS and X-ray scattering data, Eq.\,\eqref{eq: plasmon max} is adequate.
We computed the 2p2h results with the same compressibility-corrected fit of the QMC data \cite{gori2004pair} for \(S(q)\) as our GRPA values above (cf.\ \ref{app: GRPA Compressibility}).

\begin{figure}[t]
 \centering
 \includegraphics[width=0.49\textwidth]{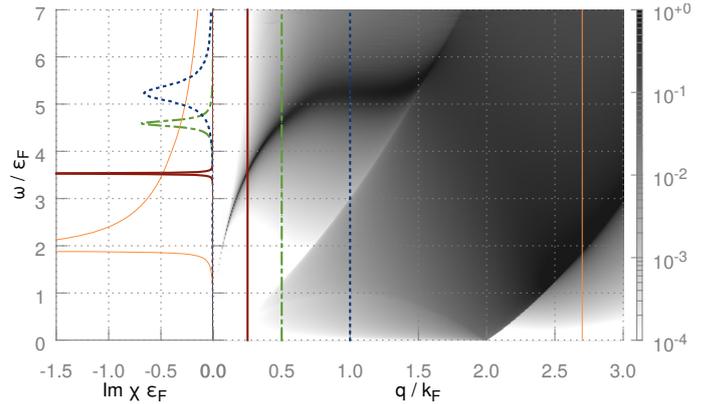}
 \caption{\label{fig: dyn response}
   Scattering loss function -Im\(\chi(q,\omega)\) (right, arbitrary
values) together with cuts at 4 characteristic wave vectors \(q_i\) (left) for a dilute 2DEL (\(r_{\scriptscriptstyle\mathrm S}=20\)) from the 2p2h theory. 
  The line styles on the left agree with those marking these \(q_i\) in the right part.
  }
\end{figure}

Figure \ref{fig: dyn response} shows the imaginary part of \(\chi^{\scriptscriptstyle\mathrm{2p2h}}\) for a highly dilute 2DEL.
Above the PHB the plasmon is visible as a strong, sharp  mode, broadened by the pair-excitations continuum.
Beyond the critical wave vector \(q_\mathrm{c}\) the mode travels, highly Landau
damped, through the PHB and regains strength near its lower edge, as is most
clearly seen in the left part of Fig.\,\ref{fig: comparedisp and qc}\,:
the rather broad orange peak is at a much lower energy than the sharp \(q\!\approx\!0.25k_{\scriptscriptstyle\mathrm{F}}\) plasmon (dark red line), 
and of much higher strength than the \(q\!=\!0.5k_{\scriptscriptstyle\mathrm{F}}\) and \(1k_{\scriptscriptstyle\mathrm{F}}\) plasmons damped by 2-pair excitations (dashed lines).

\begin{table}[b]
\footnotesize
\centering
\begin{tabular}{l|ccccc}\toprule
\(r_\mathrm{S_{\phantom|}}\)   &   2   &  5   &  10   &   20  &   30\\
\(n_{\scriptscriptstyle\mathrm{GaAs}}\) \(\left[10^{9}/\mathrm{cm}^2\right]\)    
                               &  75.2 & 12   &   3   &  0.75 & 0.33\\
\midrule\midrule
\(q_{\mathrm c}^{\scriptscriptstyle\mathrm{RPA}}\,\big[k_{\scriptscriptstyle\mathrm F}\big]\)
                               &  1.50 & 2.45 & 3.55  &  5.09 &6.28\vspace{0.08cm}\\
\(q_{\mathrm c}^{\scriptscriptstyle\mathrm{RPA}}\,[10^{5}/\mathrm{cm}]\)  &  10. & 6.8 & 4.9  & 3.5  & 2.9\\
\midrule
\(q_{\mathrm c}^{\scriptscriptstyle\mathrm{GRPA}}\,\big[k_{\scriptscriptstyle\mathrm F}\big]\), \(\,G(q,0)\)
                               & 1.13  & 1.47 & 1.70  & --    & --\vspace{0.08cm}\\
\(q_{\mathrm c}^{\scriptscriptstyle\mathrm{GRPA}}\,\big[k_{\scriptscriptstyle\mathrm F}\big]\), \(\,G_{\scriptscriptstyle\mathrm{ph}}(q)\)
                               & 1.28  & 1.54 & 1.77  &  1.93 & 2.00\\
\midrule
\(q_{\mathrm c}^{\scriptscriptstyle\mathrm{2p2h}}\,\big[k_{\scriptscriptstyle\mathrm F}\big]\)
                               &  1.05 & 1.31 & 1.44  & 1.53  & 1.55\vspace{0.08cm}\\
\(q_{\mathrm c}^{\scriptscriptstyle\mathrm{2p2h}}\,[10^{5}/\mathrm{cm}]\)
                               &  7.23 & 3.61 & 1.98  & 1.05  & 0.71\\
\bottomrule
\end{tabular}
\caption{Plasmon critical wave vector in reduced units and for a AlGaAs quantum well with areal density as given in the header.
Upper two rows: bare RPA results.  Middle two lines: Results from QMC based static local field corrections, Davoudi {\it et al.\/} \cite{davoudi01:analytical} and Eq.\,\eqref{eq: chi_GRPA Vph}.
Lower two rows: results of this work. 
}
\label{TAB: critical_wavevectors}
\end{table} 

In the static (G)RPA theories the critical wave vector \(q_\mathrm{c}\) where the plasmon hits the PHB is given by the implicit equation
\begin{equation}
 1 +\, \frac{E_{\scriptscriptstyle\mathrm F}}{V_{\!\varepsilon}(q_{\mathrm c})} \>=\>
 \sqrt{1\!+ \frac{2k_{\scriptscriptstyle\mathrm F}}{q_{\mathrm c}}} \;.
 \label{eq: qc}
\end{equation}
(Note that this also holds for layers of finite width).
The \(r_{\scriptscriptstyle\mathrm S}\!\to\!0\) (i.e.\  high density) solution is given by the RPA as \(q_\mathrm{c} \approx (2\,r_{\scriptscriptstyle\mathrm S})^{2/3\,} k_{\scriptscriptstyle\mathrm F}\). 
At intermediate densities, \(r_{\scriptscriptstyle\mathrm S\!}\approx 1\ldots5\), the two GRPA approaches and the dynamic pair theory yield comparable values, while \(q_\mathrm{c}^{\scriptscriptstyle\mathrm{RPA}}\) is markedly too high (\(>\!50\%\) at  \(r_{\scriptscriptstyle\mathrm S}\!=\!5\), see Tab.\,\ref{TAB: critical_wavevectors}).
For the highly dilute 2DELs of interest here, dynamic pair fluctuations flatten the plasmon dispersion (cf.\,Fig.\,\ref{fig: comparedisp and qc}) and, consequently, significantly  lower \(q_\mathrm{c}\) further.
For densities with \(r_{\scriptscriptstyle\mathrm S}\!\lesssim\!30\) the
numerically obtained \(q_\mathrm{c}^{\mathrm{2p2h}}\) can be accurately fitted by
\begin{equation}
 q_\mathrm{c}^{\mathrm{2p2h}}(r_{\scriptscriptstyle\mathrm S}) \>=\> 
 \frac{(2\,r_{\scriptscriptstyle\mathrm S})^{2/3} + 0.247117\,r_{\scriptscriptstyle\mathrm S} }
      {1+ 1.916638\,r_{\scriptscriptstyle\mathrm S}^{1/2} + 0.290381\,r_{\scriptscriptstyle\mathrm S} }
 \,k_{\scriptscriptstyle\mathrm F}
 \;,  \label{eq: qc_fit result}
\end{equation}
capturing both, \(r_{\scriptscriptstyle\mathrm S}\!\to\!0\)
as well as the nearly horizontal \(q_\mathrm{c}^{\mathrm{2p2h}}\approx 1.5\,k_{\scriptscriptstyle\mathrm F}\) behavior for
\(r_{\scriptscriptstyle\mathrm S}\!\gtrsim\!20\).
The comparison of the fit with the numerical results is shown in Fig.\,\ref{fig: comparedisp and qc} (dark red line and markers, respectively).

In order to facilitate comparison with experiments or other theories, we next give an approximate analytic expression for the 2p2h plasmon dispersion obtained numerically from Eq.\,\eqref{eq: plasmon max}.
Finding a formula valid for a wide range in both \(q\) and \(r_{\scriptscriptstyle\mathrm S}\) is a formidable task.
A Pad\'e inspired expression with wave vectors measured in the critical \(q_\mathrm{c}(r_{\scriptscriptstyle\mathrm S})\) given in Eq.\,\eqref{eq: qc_fit result} proved to work best.
Denoting \(\underline{q}\!\equiv\! q/q_\mathrm{c}\) and \(\omega_\mathrm{pl}^\mathrm{c}\!\equiv\omega_\mathrm{pl}(q_\mathrm{c})\) the following ansatz with the Pad\'e function of order \(n\!+\!m\) fulfills the limit \eqref{eq: kappa_SR omp} 
\begin{subequations}
  \label{eq: plasmon_fit}
 \begin{align}
  \omega^\mathrm{fit}_\mathrm{pl} \>=\> 
  \omega_\mathrm{pl}^\mathrm{c}\, \sqrt{\underline q} \>
  P_{\!_{[n,m]\!}}\big(\underline q,\, r_{\scriptscriptstyle\mathrm S}\big) 
  \;, \label{eq: plasmon_fit pade}
 \phantom{\bigg|_|}\\
  P_{[n,m]}(\underline{q}, r_\mathrm{s}) \>=\> 
  \frac{  \sum_{i=0}^n\, p_i(r_{\scriptscriptstyle\mathrm S})\,\underline{q}^i}
       {1+\sum_{j=1}^m\, \tilde{p}_j(r_{\scriptscriptstyle\mathrm S})\,\underline{q}^j}
 \;.
 \end{align}
\end{subequations}
Details on the fitting procedure and the coefficients \(p_{i}(r_{\scriptscriptstyle\mathrm S})\), \(\tilde{p}_{j}(r_{\scriptscriptstyle\mathrm S})\) are given explicitly in \ref{app: fitting_details}, Eq.\,\eqref{eq: pij_rs}, and Tab.\,\ref{tab: Meta}.
In the supplementary material we provide an implementation of our fit for several widely spread tools ({\small Origin\(^\text{\textregistered}\)},\, {\small MATLAB\(^\text{\textregistered}\)}, {\small Mathematica\(^\text{\textregistered}\)}) plus another set of coefficients, specifically suited for ultra-low densities.
 
\begin{figure}[h]
 \centering
 \includegraphics[width=0.49\textwidth]{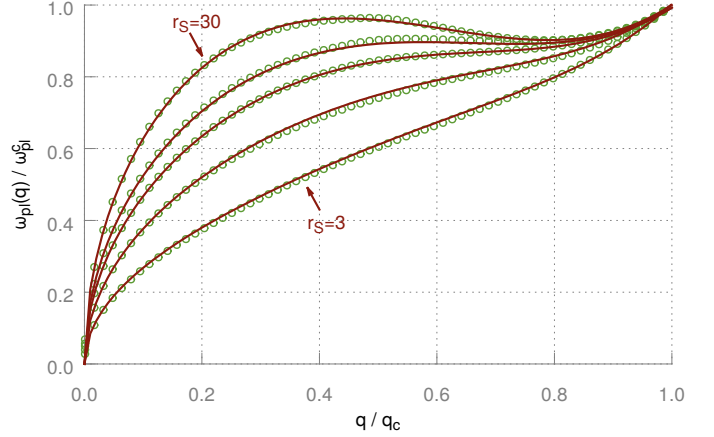}
 \caption{Plasmon dispersion from Eq.\,\eqref{eq: plasmon_fit} (lines) compared to the numerical data (points) for the full validity range of the fit, in particular,  \(r_{\scriptscriptstyle\mathrm S}\!=\! 3,\, 8.7,\, 15.2,\, 19.7, 30\). Increasing \(r_\mathrm{s}\) corresponds to higher dispersions.
The values \(r_{\scriptscriptstyle\mathrm S}=8.7,\, 15.2\) and \( 19.7\) were reported for the AlGaAs quantum well in \cite{hirjibehedin2002evidence}. 
 \label{fig: fit_data_comp} }  
\end{figure}

As seen in Fig.\ \ref{fig: fit_data_comp}, the dispersion given in Eq.\,\eqref{eq: plasmon_fit pade} accurately reproduces the numerical data over the wide density range of \(r_{\scriptscriptstyle\mathrm{S}} \!\in [3, 30]\). 
To ease comparison, all \(\omega_\mathrm{pl}(q)\) were normalized to \(\omega^\mathrm{c}_{\mathrm{pl}}\). 
For small \(q\) and in the vicinity of \(q_\mathrm{c}\) the error is well below 1\% and never exceeds 2\%.

\begin{figure}[h]
 \centering
 \includegraphics[width=0.49\textwidth]{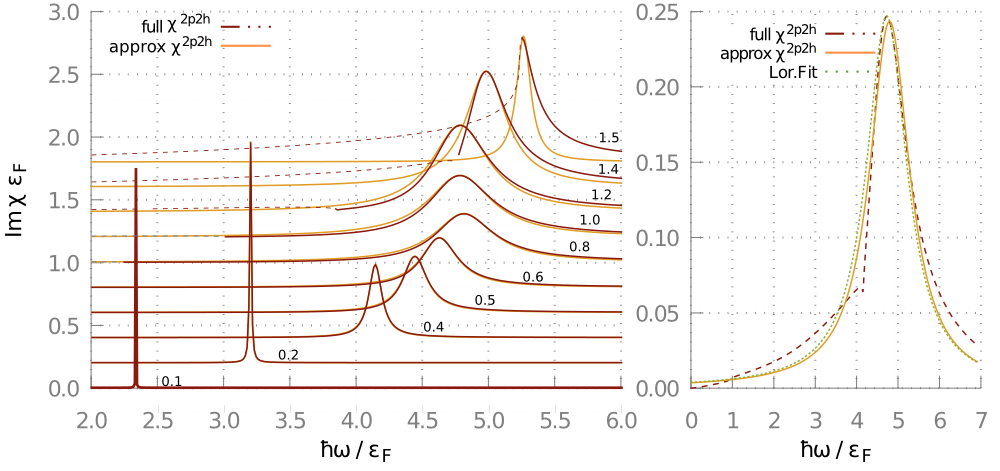}
  \caption{Full 2p2h loss function (dark red) and Lorentzian plasmon peak (orange) at \(r_{\scriptscriptstyle\mathrm{S}}\!=20\). 
    Left: Spectra for \(q/k_{\scriptscriptstyle\mathrm F}\) as indicated, shifted by 0.2\(\varepsilon_{\scriptscriptstyle\mathrm{F}}\) for better visibility. 
    Full (dashed) lines hold outside (inside) the particle--hole band (note the
asymmetry in \(\omega\) near \(q\!\approx\!q_\mathrm{c}\)).
    Right: Close up for \(q\!=\!1.3k_{\scriptscriptstyle\mathrm{F}}\!=0.85q_\mathrm{c}\): numerical (short-dashed), approximate (dotted) and fitted (full) curves.
 \label{fig: fit_Lorentzian}  }
\end{figure}

In bulk systems, multi--pair damping is negligible compared to other sources, the contribution to the life-time's dispersion, however, is significant \cite{Stur02:EEL}.
We now investigate the sheet plasmon width and the \(q-\)dependence of the 2p2h plasmon peak.
In its vicinity \(\mathrm{Im} \chi^{\scriptscriptstyle\mathrm{2p2h}}(q,\omega)\) is well represented by a Lorentzian,
\begin{equation}
  -\mathrm{Im} \chi^{\scriptscriptstyle\mathrm{2p2h}}(q,\omega) \;\approx\; 
  \frac{\Gamma_{\!\scriptscriptstyle\mathrm{2p2h}}(q)/\pi}
       {\big(\omega\!-\!\omega^{\mathrm{2p2h}}_\mathrm{pl}(q)\big)^2 \,+ \,
        \Gamma^2_{\!\scriptscriptstyle\mathrm{2p2h}}(q) }
  \,, \label{eq: 2p2hLorentz}
\end{equation}
confirmed both analyically as well as by fitting the numerically obtained \(\mathrm{Im} \chi^{\scriptscriptstyle\mathrm{2p2h}}(q,\omega)\) (see Fig.\,\ref{fig: fit_Lorentzian}).
Unless very close to the Landau damping  region, the agreement of \(\Gamma_{\!\scriptscriptstyle\mathrm{2p2h}}\) with the true FWHM is excellent.

The fit for
\(\Gamma_{\!\scriptscriptstyle\mathrm{2p2h}}(q;r_{\scriptscriptstyle\mathrm{S}})\) is given in the supplementary material (Eq.\,\eqref{S-Seq: lifetime_fit} with the coefficients of table \ref{S-STab: FWHMfit}), where we also compare the width-dispersion \(\Gamma_{\!\scriptscriptstyle\mathrm{2p2h}}(q)\) with experimental values.
Similar to the bulk, this intrinsic damping is negligible compared to that caused by `external' mechanisms (phonon and impurity scattering, inter-subband excitations, etc.). 
In contrast to 3D \cite{Stur02:EEL}, however, adding \(\Gamma(q\!=\!0)\) (either from experiment or theories beyond the electron liquid) to \(\Gamma_{\!\scriptscriptstyle\mathrm{2p2h}}(q)\), does not explain the observations here.

\section{Plasmon dispersion in semiconductor QWs \label{sec: Plasmon dispersion in semiconductor QWs}}
\subsection{Comparison with the classical dispersion \label{ssec: Comparison with the classical dispersion}}

A common method for determining the electron density from diffraction measurements is to fit the experimental plasmon dispersion to an RPA-like form. 
As discussed, the bare RPA (Eq.\,\eqref{eq: wpl_GRPA} with \(V_{\!\epsilon}(q)\!\to\! v(q)\)), underestimating correlations, grossly overestimates \(\omega_{\mathrm{pl}}(q)\).
Static correlations, further augmented by dynamic ones, act in the opposite direction.
Temperature effects raise \(\omega_{\mathrm{pl}}(q)\), while increasing the layer width softens it \cite{HwaS01:plasmon,fukuda2007plasmon}:
The smeared out wave function \(\varphi_{\!\scriptscriptstyle0}(z)\) of the quantum well reduces the effective interaction and thus lowers the RPA correlations (approaching the bulk result for very wide wells would require to account for multiple subbands).

Obviously, comparing theory with measurements would require a precise, independent experimental determination of more parameters than possible.

Specifically, \(r_{\scriptscriptstyle\mathrm S}\) (\textit{i.e.} the areal electron density \(n\)) is subject to some ambiguity \cite{HwaS01:plasmon}. In \cite{hirjibehedin2002evidence} it was determined from fitting the plasmon dispersion to the empirical form
\begin{equation}
 \omega^{\mathrm{exp}}_\mathrm{pl}(q) \>=\> 
 \omega_0(q)\, \sqrt{1+q\,\xi\,}  
 \;, \label{eq: xi_approxi empircal}
\end{equation}
where the length \(\xi\) contains all effects due to temperature \(T\), well-width \(L\), and correlations; 
the latter, in turn, are split into RPA (\(\equiv\) `non-local') and LFC contributions,
\begin{subequations}
\label{eq: xiology}
 \begin{align} 
  \xi\!\equiv \xi^{L,T}_{\,\mathrm{tot}}
   &\equiv\> \xi^{0,0}_\mathrm{tot}  +\xi^{0,T} - \xi^{L,0}
  \;,\label{eq: baptizing xis}
  \\
  \xi^{0,0}_{\mathrm{tot}} &=
    \big(\xi^{\scriptscriptstyle\mathrm{RPA}}_{\mathrm{cor}} -
         \xi^{\scriptscriptstyle\mathrm{LFC}}_{\mathrm{cor}}\big)^{\!0,0} \,\equiv\>
   +\xi_{\mathrm{nloc}}^{0,0} -\xi_{\mathrm{cor}}^{0,0} 
  \;. \label{eq: xi_splitting}
 \end{align}
\end{subequations}
(see \ref{app: Plasmon dispersion coefficients} about terminology).
For very low temperatures the observed \(\xi\) scales with \(k_{\scriptscriptstyle\mathrm F}^{-1}\); 
since in the 2DEG \(\xi_{\mathrm{nloc}}\) is density--independent, the zero temperature limit is interpreted \cite{hirjibehedin2002evidence} as the correlation part
\begin{equation}
 \xi_{\mathrm{exp}} \xrightarrow[\to0]{T}\> -(0.17\!\pm\!0.04)\,\sqrt{2}/k_{\scriptscriptstyle\mathrm F}
 \>\equiv\> -\xi^{L,0}_{\,\mathrm{cor}}(r_{\scriptscriptstyle\mathrm S}) \Big|_\mathrm{exp}
 \;. \label{eq: xi Hirji}
\end{equation}

\begin{figure}[h]
 \centering
 \includegraphics[width=0.49\textwidth]{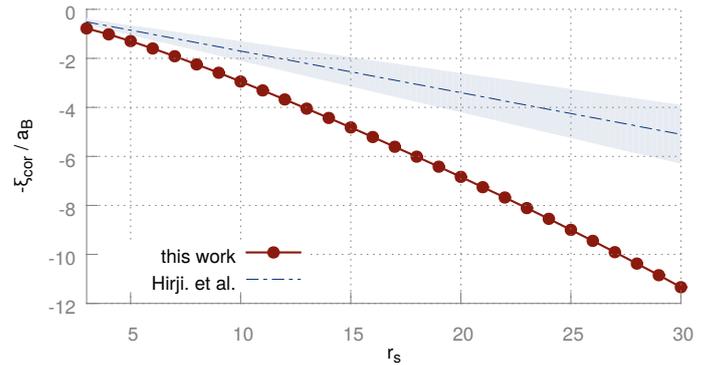}
 \caption{Plasmon correlation wavelength \(\xi^{\scriptscriptstyle0,0}_\mathrm{cor}\) from the 2p2h theory for a strictly 2DEL (solid line: fit \eqref{eq: plasmon_fit}, markers: numerical data) and measured estimate for a finite width AlGaAs 2DEL \cite{hirjibehedin2002evidence} (double-dashed line with shaded error bar).
  \label{fig: xi_comp} 
}
\end{figure}

The RPA term \(\xi^{\scriptscriptstyle 0,0}_{\mathrm{nloc}}= 3a_{\scriptscriptstyle\rm B}^*/8\) follows from Eq.\,\eqref{eq: wpl_GRPA qto0} with \(V_{\!\epsilon}(q)\!\to\! v(q)\).
From the small \(q\) expansion of Eq.\,\eqref{eq: plasmon_fit}) we here provide a state-of-the-art result for the correlation coefficient due to two-pair excitations in the strictly 2DEL:
\begin{equation}
 -\xi^{0,0}_{\,\mathrm{cor}}(r_{\scriptscriptstyle\mathrm S})\Big|_\mathrm{2p2h} \>=\> 
  \frac2{q_\mathrm{c}(r_{\scriptscriptstyle\mathrm S})}\,\Big(\frac{p_1}{p_0} - \tilde{p}_1 \Big)
 \,- \frac38\,a_{\scriptscriptstyle\rm B}^*
 \;. \label{eq: xi fit}
\end{equation}
In Fig.\,\ref{fig: xi_comp}\, this is compared with \(\xi^{L,0}_{\,\mathrm{cor}}\) determined from Eq.\,\eqref{eq: xi Hirji}.
As expected, the computed strictly 2D correlation effects are larger in magnitude than those measured for  \(L\!\approx330\)\AA.

In a QW with lowest subband wave function \(\varphi_{\!\scriptscriptstyle0}\) the 3D density \(\rho({\bf r},z)\)  can often be approximated as \(n\,|\varphi_{\!\scriptscriptstyle0}(z)|^2\).
The 2D Coulomb potential is then modified with an \(r_{\scriptscriptstyle\mathrm S}-\)independent `form factor' \(F(q)\),%
\begin{subequations}
 \label{eq: vwidth}
 \begin{align}
   v(q) &\to v(q)\,F(q) \\
   F(q) &= \int\!\!dz_1\! \int\!\!dz_2\> \big|\varphi_{_{0\!}}(z_1)\big|^2 \, e^{-q|z_1-z_2|}\,\big|\varphi_{_{0\!}}(z_2)\big|^2
   \;.  \label{eq: Fwidth}
 \end{align}
\end{subequations}
Clearly, this yields a density-independent dispersion coefficient \(\xi^{L,0}\), where
\begin{subequations}
 \begin{align}
   F(q\!\to\!0) &=   1 - q\,\xi^{L,0}
   \;;\quad  \xi^{L,0}_{} \!>\!0 
   \phantom{\Big|_|}\\
   \omega^{L,0\!}_\mathrm{pl}(q\!\to\!0)\Big|_{\scriptscriptstyle\mathrm{RPA}} &=\>
   \omega_0(q)\, \sqrt{1+ \big(\textstyle\frac38\displaystyle a_{\scriptscriptstyle\rm B}^* - \xi^{L,0}\big)\,q\,}
   \;. \label{eq: wpl_RPAL qto0xi}
 \end{align}
\end{subequations}
The full RPA finite \(L\) plasmon is given by \(V_{\!\epsilon}\to vF\) in Eq.\,\eqref{eq: wpl_GRPA qto0} (cf.\ \ref{app: Bare RPA finite width dispersion} for details).

While both these contributions to the plasmon dispersion are constant, the LFC part must increase with \(r_{\scriptscriptstyle\mathrm S}\).
Since correlations are the stronger the thinner the QW and/or the higher \(r_{\scriptscriptstyle\mathrm S}\), 
the discrepancy in Fig.\,\ref{fig: xi_comp} increases for dilute systems.
Cum grano salis, the computed \(L\!=\!0,\, T\!=\!0\) data presented in the figure can thus be considered as a lower bound for measurements.

We next turn to incorporating finite \(L\) and finite \(T\) effects into our approach. 
The static GRPA theories of Sec.\,\ref{sec: GRPA} both rely on equilibrium QMC results, one needing \(G(q,\omega;\,L\!=\!0,T\!=\!0)\), the other \(S(q;\,L\!=\!0,T\!=\!0)\). The latter function is also input to the 2p2h theory.
No QMC results for \(G(q,0;L,T)\) and \(S(q;L,T)\) are available.
For the plasmon we therefore adopt the above intuitive strategy, 
\begin{equation}
  \omega^{L,T\!}_\mathrm{pl}(q;r_{\scriptscriptstyle\mathrm S})
  \>=\>
  \omega^{0,0\!}_\mathrm{pl}(q,r_{\scriptscriptstyle\mathrm S})
       \,\sqrt{1+\big(\xi^{0,T\!}(r_{\scriptscriptstyle\mathrm S}) \!-\! \xi^{L,0}\big)\,q \,}
  .\; \label{eq: plasmon_fit_temp_width}
\end{equation}
The parameters \(\xi^{\scriptscriptstyle 0,T}\) and \(\xi^{\scriptscriptstyle L,0}\) are taken from the experiment under consideration, \(\omega^{0,0\!}_\mathrm{pl}\) from the fit \eqref{eq: plasmon_fit} of the 2p2h result.
For long wavelengths \eqref{eq: plasmon_fit_temp_width} is exact and identical to
\begin{equation}
  \omega_{0}\, \sqrt{1+ \big(\xi^{0,0\!}_{\mathrm{tot}}(r_{\scriptscriptstyle\mathrm S}) \!+\!
                             \xi^{0,T}(r_{\scriptscriptstyle\mathrm S}) \!-\! \xi^{L,0}\big)\,q \,}
  ,\; 
\end{equation}
(with the classical \(\omega_{0}\propto\! \sqrt{q}/r_{\scriptscriptstyle\mathrm S}\) and notation as in Eq.\,\eqref{eq: xiology}).

\subsection{Application: electron density \label{ssec: Application: electron density}}

\begin{figure}[h]
 \centering
 \includegraphics[width=0.49\textwidth]{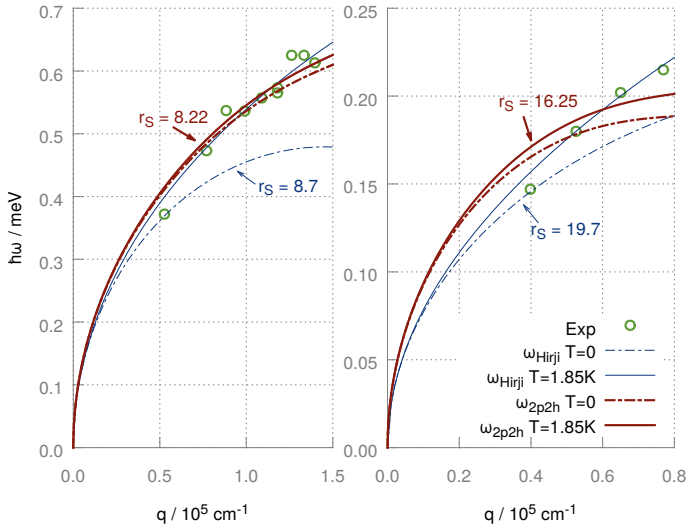}
 \caption{Measured plasmon energies (green dots) for two samples, fitted to Eq.\,\eqref{eq: xi_approxi empircal}, including both, finite temperature and finite width dispersion (blue solid lines), and for \(\xi^{\scriptscriptstyle0,T\!}\!=\!0\!=\!\xi^{\scriptscriptstyle L,0}\) (dashed--dotted blue lines) \cite{hirjibehedin2002evidence}.
  Dark-red dispersions: fit of the theoretical 2p2h results \eqref{eq: plasmon_fit_temp_width} to the experimental data, 
  again with and without \(\xi^{\scriptscriptstyle0,T\!},\,\xi^{\scriptscriptstyle L,0}\) (solid and dashed--dotted lines, respectively).
 \label{fig: pldisp comp Hirji}}
\end{figure}

We apply this to the \(L\!=\!330\)\AA\ GaAs QW experimentally studied at \(T\!=1.85K\) in \cite{hirjibehedin2002evidence,EPDH00:collective}.
There, two samples were fitted to the empirical form (\ref{eq: xi_approxi empircal}), resulting in 
\(r_{\scriptscriptstyle\mathrm S}\!=8.7\) and \(19.7\), respectively (full thin blue lines in Fig.\,\ref{fig: pldisp comp Hirji}).
Taking  \(\xi^{0,T}\) from Fig.\,4 of \cite{hirjibehedin2002evidence}, a least square fit of \eqref{eq: plasmon_fit_temp_width} yields the somewhat different values \(r_{\scriptscriptstyle\mathrm S}\!=8.22\) (\(n\approx 4.46\!\times\! 10^9\mathrm{cm^{-2}}\)) and \(16.25\) (\(n\approx 1.14\!\times\! 10^9 \mathrm{cm^{-2}}\)) (full thick red lines in Fig.\,\ref{fig: pldisp comp Hirji}).
For the denser sample (left panel), theory and experiment agree nicely; with an \(r_{\scriptscriptstyle\mathrm S}-\)difference of \(\sim\!5\)\%.

The ultra-dilute case (where \(r_{\scriptscriptstyle\mathrm{S}}\) differs by 15\%) is less satisfactory, the 2p2h curve being too flat compared to the experiment.
Conversely, in 2D \(^3\)He \cite{godfrin12:observation} dynamic pair correlations proved crucial to explain the measured spectrum.
At a density of \(1/a^2\pi\!\approx\!10^9\,\mathrm{cm}^{-2}\), the plasmon with \(q\!=0.7\!\times\!10^5\,\mathrm{cm}^{-1}\) has a wavelength of \(\lambda\!=2\pi/q\!\approx\!5a\) and thus definitely should `feel' the two--body fluctuations.
We attribute the discrepancy to the fact that the width \(L\) not only diminishes the RPA \(q\!\to\!0\) dispersion, but reduces correlations in general.  
Since larger \(q\) are more affected by correlation effects, their reduction will also be larger there, diminishing the negative curvature of \(\omega^{\scriptscriptstyle\mathrm{2p2h}\!}_\mathrm{pl}(q)\)
(see \ref{app: Bare RPA finite width dispersion} for further details).

\section{Conclusions \label{sec: Conclusions}}

We calculated the excitation spectrum of the 2DEL including 2p2h excitations, with special emphasis on the application to AlGaAs QWs.
We found that dynamic pair correlations lower the sheet plasmon's dispersion massively.
The agreement with experiments is good, except for ultra-dilute sytems, where finite width effects should be better accounted for. 
This requires the availability of high-quality data for the \(L\!\ne\!0\) pair distribution function; work in this direction is in progress.
For zero well width we provided a fit of the plasmon dispersion
\(\omega^{\scriptscriptstyle\mathrm{2p2h}\!}_\mathrm{pl}(q;r_{\scriptscriptstyle\mathrm{S}})\), valid in the whole range of \(r_{\scriptscriptstyle\mathrm{S}}\!\in [3,\ldots,30]\) and \(q\in [0,\ldots,q_\mathrm{c}]\) (see download in the Supplementary Material).

An extension of the dynamic pair theory to spin-dependent effective dynamic interactions in partially spin--polarized 2DELs \cite{agarwal2014long} appears of practical interest, due to the low--loss attribute of the magnetic antiresonance \cite{kreil2015excitations}.
The input functions needed for such systems are available for \(L\!=\!0\) \cite{gori2004pair}
and currently studied in our group for \(L\!\ne\!0\). Another topic worth pursuing is the influence of dynamic correlations on the effective mass enhancement \cite{ADPG05:quasiparticle, AGTP09:effectivemass, KroS03:physical}.

\section*{Acknowledgements}

We thank Margherita Matzer and Johanna Herr for valuable assistance with  {\small Origin\(^\text{\textregistered}\)} and {\small MATLAB\(^\text{\textregistered}\)}.
DK acknowledges financial support by the Wilhelm Macke Stipendienstiftung and the Upper Austrian government (Inovatives Ober\"osterreich 2020).


\begin{appendix}

\section{2-Pair Fluctuations}
\label{app: 2-Pair Fluctuations}

Our approach extends the widely used Jastrow-Feenberg ansatz for a many--particle ground state to excited states \cite{Feenberg69:theory, ChaC76:density}.
Both share the advantage of avoiding the summation of many, mutually cancelling diagrams by obtaining the correlations via an optimization procedure.

The ground state wave function is approximated as 
\begin{equation}
    |\Psi_{\rm g} \rangle  \>=\> 
    \textstyle\frac1{{\cal I}^{1/2}_{\!\rm g}}\; \displaystyle
    \exp\!\big(\overline U_{\!\rm g}\big) \> 
    | \Phi_{\!_0} \rangle \;,
 \label{eq: Jastrow WaveFunction}
\end{equation}
(\({\cal I}_{\!\rm g}\) is the normalization integral, \(\Phi_{\!_0}\) a Slater--determinant).
The correlation operator \cite{Feenberg69:theory} \(\,\mathrm{exp}(\overline U_{\!\rm g})\) invokes n-body equilibrium correlation functions \(\bar u_n({\bf r}_1,\ldots,{\bf r}_n)\) 
\begin{equation}
  \overline U_{\!\rm g} \equiv\>
    \textstyle\frac12\displaystyle\!\sum_{\scriptscriptstyle i<j}
    \bar u_2(|{\bf r}_i\!-\!{\bf r}_j|) \>+
    \textstyle\frac12\displaystyle\!\!
    \sum_{\scriptscriptstyle i<j<k\!}\bar u_3({\bf r}_i,{\bf r}_j,{\bf r}_k) \>+\> \ldots  \;,
  \label{eq: JastrowCorrelations} 
\end{equation}
obtained optimally from minimizing the energy \(E_{\rm g}\). 
This focuses right away on the comparably small correlations, avoiding the summation of large perturbational terms with opposite sign.

The {\it Dynamic Many Body Theory\/} \cite{CaKL15:dynamic, BHKP10:dynamic, godfrin12:observation, ChaC76:density}
generalizes this idea to a system perturbed by \(H_{\rm ext}(t)\) (the full Hamiltonian is
\(\, \widehat H_{\rm tot}\! = \widehat H \!+\! \widehat H_{\rm ext}\)).
The perturbed wave function takes a form analogous to Eq.\,\eqref{eq: Jastrow WaveFunction}:
\begin{equation}
 |\Psi(t)\rangle \>=\>
    \textstyle \frac1{{\cal I}^{1/2}_{\! t}} \displaystyle\; 
    \exp\!\big(\overline U_{\!\rm g}\big) \> 
    \exp\!\big(\widehat U_t\big) \> 
    | \Phi_{\!_0} \rangle \;
 \label{eq: psitime}
\end{equation}
(again, \({\cal I}_{t}\) ensures the normalization).
We abbreviate \(k\!\equiv ({\bf k},\sigma_{\!\scriptscriptstyle k})\) for wave vector and spin.
The excitation operator reads
\begin{eqnarray}
   \widehat U_t &\equiv&  
   - \textstyle\frac{it}{\hbar}\displaystyle E_{\rm g} \,
   + \sum_{\scriptscriptstyle p_1h_1} u^{(1)}_{p_1h_1\!}(t)\;
   a^\dagger_{p_1\!} a^\qfd_{h_1}
  \label{eq: Uop} \\
    &+& \sum_{\scriptscriptstyle p_1\!h_1p_2h_2}\!  \textstyle{1\over 2}\displaystyle
          u^{(2)}_{p_1h_1p_2h_2\!} (t)\; 
   a^\dagger_{p_1\!}a^\qfd_{h_1}\, a^\dagger_{p_2\!} a^\qfd_{h_2} 
   \phantom{\bigg|^|} \;.  \nonumber
\end{eqnarray}
It creates $n$-particle\,--\,$n$-hole excitations (\(n\!=\!1,2\)) from the free determinant \( |\Phi_{\!_0}\rangle\), dynamically coupled by the ``$n$-pair--fluctuations'' \(u^{\scriptscriptstyle(n)}_{\scriptscriptstyle p_1\ldots h_n}\),
finally correlated by $\overline U_{\!\rm g}$.
The fluctuations are again determined via functional optimization (`least' 
action principle).

The deviations \(\delta A\! \equiv  \langle\widehat A \!-\! A_{\rm g}\rangle\) of an observable \(\widehat A\) from its unperturbed value \(A_{\rm g}\) are now calculated with the wave function \eqref{eq: psitime}.
In linear response, only first order terms in \(u^{(n)}\) need to be kept. This results in a sum over \(n-\)pair fluctuations weighted with matrix elements,
\begin{equation}
 \delta A =\> {\Re}e \sum_{n=1}^2 
 \textstyle\frac1{n!}\displaystyle\!\!
 \sum_{p_1\ldots h_n} 
 \overline{\cal A}_{0,\,p_1\ldots h_n} \>
 u^{(n)}_{p_1\ldots h_n} \,.
 \label{eq: Acbfn}
\end{equation}
The transition integrals \(\overline{\cal A}_{0,\,p_1\ldots h_n}\) involve, due to \(\delta \widehat U\) in \eqref{eq: Uop}, \(n-\)pair excited states,
\begin{equation}
  | \Phi_{p_1\ldots h_n} \rangle \equiv\> 
    \exp\!\big(\overline U_{\!\rm g}\big) \> 
    a^\dagger_{p_1\!}a^\qfd_{h_1}\! \ldots a^\dagger_{p_n\!} a^\qfd_{h_n} \,
    | \Phi_{\!_0} \rangle \;.
  \label{eq: States}
\end{equation}
These form a complete but not orthogonal set (\textit{c.f.\ }the ``correlated basis functions'' in \cite{KroTriesteBook}). 
We denote the plain overlaps, \textit{i.e.\ }those of \(\delta\widehat A\!=\!1\), as \(\overline{\cal N}_{\!0,\,p_1\ldots h_n}\).
Evaluating these high-dimensional integrals is rather uneconomical.
Much more promising is a localization strategy.

A prime example involves the static structure factor,
\begin{equation}
 S(q) \>\equiv\> \langle \delta\widehat\rho_{\bf q}\, 
                         \delta\widehat\rho_{-\bf q} \rangle/N
 \;.
\end{equation}
Expressing the density fluctuations via creation-- and annihilation operators and applying \(\delta\widehat\rho_{\bf q}\delta\widehat\rho_{-\bf q}\) to \(|\Psi_{\!\rm g}\rangle\) leads to
\begin{equation}
 S(q) \>=\> S^0(q) +\> \textstyle\frac1N\displaystyle\!\sum_{h_1h_2}
 \overline{\cal N}_{\!0,\,p_1h_1p_2h_2} \,,\; \mbox{\footnotesize\(
 \begin{array}{ccc} 
   {\bf p_1}\equiv {\bf h_1}\!-\!{\bf q} \\ 
   {\bf p_2}\equiv {\bf h_2}\!+\!{\bf q} \end{array}\)}
 \;,
 \label{eq: SqN2pair}
\end{equation}
with the free structure factor \(S^0(q)\) \cite{giuliani2005quantum}.
Knowledge of \(S(q)\) from any state-of-the-art theory can therefore be used to replace the \(h_{i}-\)summed \(\overline{\cal N}_{\!0,\,p_1\ldots h_2}\,\).
This example captures the idea: Unknown {\it non-local\/} matrix elements \(\,\overline{\cal A}\,\) are approximated by known {\it local\/} functions \({\cal A}_{\scriptscriptstyle\rm av}^{\scriptscriptstyle(\ell,n)}\), depending only on the transferred momenta:
\begin{equation}
  \overline{\cal A}_{p'_1\ldots h'_\ell,\,
               p_1^{\phantom{,\!}}\ldots h_n^{\phantom{,\!}}} 
  \>\longrightarrow\;
  {\cal A}_{\scriptscriptstyle\rm av}^{\scriptscriptstyle(\ell,n)}
          (\mbox{\small\({\bf q}'_1,\ldots,{\bf q}_{n-1}^{\phantom{,\!}}\)}) 
 \,,\;\mbox{\footnotesize\(\begin{array}{ccc} 
   {\bf q_i}\equiv {\bf p_i}\!-\!{\bf h_i} \\ 
   {\bf q_i'}\equiv {\bf p_i'}\!-\!{\bf h_i'} \end{array}\)}
 \qquad \label{eq: FSav strategy}
\end{equation}
(momentum conservation implies that the sum of all \({\bf q}_i\) equals that of all \({\bf q}'_i\)).  
For the ground state quantities \({\cal A}_{\scriptscriptstyle\rm av}^{\scriptscriptstyle(\ell,n)}\) the best available data (e.g.\ \(S(q)\) from QMC simulations \cite{gori2004pair}) are taken.

The density response follows from Eq.\,\eqref{eq: Acbfn} as
\begin{eqnarray}
 \hspace{-0.3cm}
 \delta \rho({\bf q};t) &=& {\Re}e \bigg\{
   \sum_{p_1'\ldots h_1} \overline{\cal N}_{\!p'_1h'_1,\,p_1^{\phantom{'}}h_1^{\phantom{'}}} 
       \> u^{(1)}_{p_1h_1\!}(t) \>+\>
   \label{eq: rhocbf}
   \\\nonumber&&\hspace{0.0cm}
   \sum_{p_1'\ldots h_2} 
  \overline{\cal N}_{\!p'_1h'_1,\,p_1^{\phantom{'}}h_1^{\phantom{'}}p_2^{\phantom{'}}h_2^{\phantom{'}}} 
       \> \textstyle\frac12\displaystyle u^{(2)}_{p_1h_1p_2h_2\!}(t) \bigg\}\;.
\end{eqnarray}
The local approximations of the \(\overline{\cal N}-\)matrices are all closely related to the ground state \(n-\)particle structure factors \cite{BHKP10:dynamic}.
Next, the fluctuation amplitudes \(u^{(1)},\,u^{(2)}\) are determined from Euler--Lagrange equations (EL-eqs).

The Lagrangian corresponding to Schr\"odinger's equation and the ansatz \eqref{eq: psitime} give
\begin{subequations}
 \begin{eqnarray}
  {\cal L}(t) =\> \langle\Psi(t)\big|\, \widehat H_{\rm tot}\!-
    i\hbar\frac{\partial}{\partial t}\, \big|\Psi(t)\rangle \;,
  \label{eq: Lagr}
   \\
   \bigg(\, \mbox{\small\(\displaystyle 
      \frac{\delta}{\delta u^{*(n)}_{p_1\ldots h_n}} \,-\, \frac{d}{dt}
      \frac{\delta}{\delta \dot u^{*(n)}_{p_1\ldots h_n}} \)}
   \,\bigg)\, {\cal L}(t) \>=\> 0
  \;. \label{eq: EL eqs allgemein}
 \end{eqnarray}
 \label{eq: EL start}
\end{subequations}
For excitation operators of the type \eqref{eq: Uop} the time derivative
term yields for the lhs of \eqref{eq: EL eqs allgemein}
\begin{equation}
 \textstyle\frac{i\hbar}{{\cal I}_{\!t}}\displaystyle 
 \big\langle
  e^{\overline{U}_{\!\mathrm{g}}} e^{\widehat U^\dagger_t}
  a^\dagger_{p_1}\!\ldots
  a^{\phantom\dagger}_{h_n}\! \Phi_{\!_0}\, \big|
  \big(1\!-\! |\Psi\rangle\langle\Psi|\big) \big|\,
  e^{\overline{U}_{\!\mathrm{g}}} e^{\widehat U^\dagger_t}\,
  \widehat{\dot U}_{\!t\,} \Phi_{\!_0} \big\rangle .
\end{equation}
In linear response this invokes the \(\dot u^{(n)}_{p_1\ldots h_n}\) and the \(\overline{\cal N}\)-matrix elements of Eq.\,\eqref{eq: rhocbf}. 
Due to
\begin{equation}
 \widehat H_{\rm ext} = \int\!\!d^2r\> V_{\!\rm ext}({\bf r},t)\,
 \widehat\rho({\bf r}) \;,
\end{equation}
these also enter the perturbation contribution of \eqref{eq: EL start}.

The remaining, essential parts of the EL-eqs arise from \(\langle \Psi(t)|\widehat H|\Psi(t)\rangle\).
We first take the functional derivative and then calculate the
expectation value in linear response as outlined above, now for the operator
\(\widehat A \to a^\dagger_{p_1\!}\ldots a_{h_n}^{\phantom{'}} \widehat H\).
This brings the transition integrals of the Hamiltonian into play:
\begin{equation}
 \overline{\cal H}_{p_1^{\phantom{'}}\ldots h_n^{\phantom{'}},p_1'\ldots h_n'}  \quad\mbox{and}\quad
 \overline{\cal H}_{0,\,p_1\ldots h_n^{\phantom{'}}}
 \;. \label{eq: Hcbfn}
\end{equation}
For \(\widehat H_{\mathrm{ext}}\!=\!0\) the EL-eqs must be fulfilled with \(\widehat U_t\!=\!0\) (equilibrium condition). 
From this the optimal local functions \({\cal H}_{\scriptscriptstyle\rm av}^{\scriptscriptstyle(\ell,n)}\) are determined in the spirit discussed above.
With \(t_k\!\equiv \hbar^2k^2/2m\), end up with the diagonal and off-diagonal
Hamiltonian matrix elements approximated as
\begin{equation}\begin{array}{rlllll}
  \displaystyle \overline{\cal H}_{p_1\ldots h_n,p_1\ldots h_n} 
  \!\rightarrow&\!\!\! \sum_{i=1}^n \big(t_{p_i}\! - t_{h_i}\big)
    \equiv \sum_{i=1}^n e_{p_i,h_i} \;,
 \vspace{0.3cm}\\\displaystyle
  \overline{\cal H}_{p_1^{\phantom{'}}\ldots h_n^{\phantom{'}},p_1'\ldots h_n'} 
  \!\rightarrow&\!\!\!
    \!\frac12\!\sum_{i=1}^{\scriptscriptstyle n+n'\!}
    \big( e_{p_i,h_i}\! - \textstyle\frac{t_{q_i}}{S^0(q_i)}\displaystyle \big)\, 
    {\cal N}_{\scriptscriptstyle\rm av\!}
    (\mbox{\footnotesize \({\bf q_1},\ldots,{\bf q_n'}\)}) \,,
 \end{array}
 \label{eq: Hcbf d_and_off}
\end{equation}
respectively.
Here, the 2p2h-part \(\overline{\cal H}_{p_1^{\phantom{'}}\ldots h_2^{\phantom{'}},p_1'\ldots h_2'}\)
needs the 4--particle structure factor, which we take in a product approximation.
For details beyond these key steps of the derivation, we refer to \cite{BHKP10:dynamic}.

\section{Collective Approximation}
\label{app: Collective Approximation}

Valuable insight on the 2p2h expressions is gained from their PPA forms.
Using the Bijl-Feynman energies \(\varepsilon_q^{\scriptscriptstyle\mathrm{BF}}\equiv \hbar^2q^2/2mS(q)\) and the PPA partial Lindhard functions,
\begin{equation}
 \chi^{0\scriptscriptstyle\pm\!}_{\!\scriptscriptstyle\mathrm{PPA}\!}(q,\omega) \>=\> 
 \frac{\pm S^0(q)}{\hbar\omega \mp \frac1{S^0(q)}\frac{\hbar^2q^2}{2m} +i0^\pm}
 \label{eq: Chi0pmCA}
\end{equation}
immediately yields
\begin{equation}
 \varkappa^{\scriptscriptstyle\pm\!}_{\scriptscriptstyle\mathrm{PPA}\!}(q,\omega) \>=\> 
 \frac{\pm S(q)}{\hbar\omega \mp \varepsilon_q^{\scriptscriptstyle\mathrm{BF}} +i0^\pm}
 \;. \label{eq: kapppmCA}
\end{equation}
Obviously, \(\varkappa^{\scriptscriptstyle+}_{\scriptscriptstyle\mathrm{PPA}} \!+
             \varkappa^{\scriptscriptstyle-}_{\scriptscriptstyle\mathrm{PPA}} \)
coincides with \(\chi^{\scriptscriptstyle\mathrm{GRPA}}_{\scriptscriptstyle\mathrm{PPA}}\) with
 \(V_{\!\epsilon}\!=\!V^0_{\!\scriptscriptstyle\mathrm{ph}} \,\) from Eq.\,\eqref{eq: Vph0}.
The PPA polarizability \(\Pi_{\mathrm{s}}\) reads
\begin{equation}
  \Pi_{\mathrm{s,\scriptscriptstyle PPA}}^{\scriptscriptstyle\pm} \>=\>
   \frac{\pm S^0}{\hbar\omega \mp \frac1{S^0}\,\big[\frac{\hbar^2q^2}{2m} +
    W^{\scriptscriptstyle\pm}_{\mathrm{s}} \big] +i0^\pm} \;.
 \label{eq: PisCA}
\end{equation}
These simpliefied functions imply a boson-like 2p2h density response:
\begin{subequations}
 \begin{align}
  \chi^{\!^{\mathrm{2p2h}\!}}_{\!\scriptscriptstyle\mathrm{PPA}}(q,\omega) \>=\;&
   + \frac{S(q)}{\hbar\omega - \varepsilon_q^{\scriptscriptstyle\mathrm{BF}} 
              - \frac{1}{S(q)} W^{\scriptscriptstyle+\!}_{\phantom{s}}(q,\omega) +\!i0^+}
  \nonumber\phantom{\Bigg|_{\big|}}\\ &
  - \frac{S(q)}{\hbar\omega + \varepsilon_q^{\scriptscriptstyle\mathrm{BF}} 
              + \frac{1}{S(q)} W^{\scriptscriptstyle-\!}_{\phantom{s}}(q,\omega) +\!i0^+}
 \;,\end{align}
 where \(W^{\scriptscriptstyle\pm}_{\phantom{s}}\) is calculated with the PPA pair propagator
 \begin{equation}
  \mu^{\scriptscriptstyle\pm\!}_{\!\scriptscriptstyle\mathrm{PPA}}(q',q'',\omega) \>=\>
  \frac{\pm\,S(q')\,S(q'')}
       {\hbar\omega \mp \varepsilon_{q'} ^{\scriptscriptstyle\mathrm{BF}} 
                    \mp \varepsilon_{q''}^{\scriptscriptstyle\mathrm{BF}} \,+\!i0^+}
 \;. \label{eq: muCA}
\end{equation}
\end{subequations}
Neglecting triplet correlations in the vertex \eqref{eq: vertexK} results in
\begin{align}
  W^{\scriptscriptstyle\pm\!}(q,\omega) \>=\> 
  \textstyle\frac1N\!\!\sum\limits_{\mathbf{q}'} \displaystyle
  \textstyle\frac{\hbar^2\mathbf{q}\cdot\mathbf{q}'}{2m}\displaystyle
  X(q')&\big[\,
  \textstyle\frac{\hbar^2\mathbf{q}\cdot\mathbf{q}'}{2m}\displaystyle
  X(q')  
  \\\nonumber &\,+\,
  \textstyle\frac{\hbar^2\mathbf{q}\cdot\mathbf{q}''}{2m}\displaystyle
  X(q'')  \big]
   \mu^{\scriptscriptstyle\pm\!}(q',q'',\omega) 
  \;. \label{eq: Wpm nou3}
\end{align}

\section{GRPA Compressibility}
\label{app: GRPA Compressibility}

For the response function \eqref{eq: def chi G} with a static local field correction 
\begin{equation}
 \chi(q, \omega) = \frac{\chi^0(q,\omega)}{1-v(q)\,(1\!-\!G(q))\,\chi^0(q, \omega)} 
\end{equation}
the compressibility sum rule implies the condition
\begin{equation}
 G(q\!\rightarrow\!0) = \frac12\left(1-\frac{\kappa^0}{\kappa} \right) q a_{\scriptscriptstyle\mathrm B}^*
 \;.
\end{equation}
The high frequency limit of the Lindhard function 
\begin{equation}
\chi^0(q\!\to\!0,\omega) \approx \frac{q^2}{m \omega^2} \left[ 1 + \frac{3}{4} \frac{q^2 v_\mathrm{F}^2}{\omega^2}\right]
\end{equation}
then leads to the long wavelength plasmon dispersion
\begin{equation}
 \omega^{\scriptscriptstyle\mathrm{GRPA}}_\mathrm{pl}(q\!\to\!0) \>=\> \omega_0(q)
 \left[1 + \frac{1+2\kappa/\kappa^0}{2^{5/2}\,r_{\scriptscriptstyle\mathrm S}} \frac{q}{k_{\scriptscriptstyle\mathrm F}}
 + \mathcal{O}\big(q^2\big) \right] \;.
 \;\label{eq: long_wavelength_omega_pl}
\end{equation} 
For the dynamic local field factor 
\begin{equation}
  v(q)\,G_{\scriptscriptstyle\mathrm{2p2h}}(q, \omega) \,=\, 
  \frac1{\chi^{\scriptscriptstyle\mathrm{2p2h}}(q, \omega)} - \frac{1}{\chi^0(q, \omega)} + v(q) 
  \\\phantom{\Bigg|_|}
\end{equation}
the long wavelength limit
\begin{equation}
  G_{\scriptscriptstyle\mathrm{2p2h}}(q\!\to\!0, \omega)  \,=\, 1- \frac{V_{\scriptscriptstyle\mathrm{ph}}(q)}{v(q)}
  \,=\, G_{\scriptscriptstyle\mathrm{ph}}(q) 
\end{equation}  
coincides with its static counterpart.
Therefore, the plasmon dispersion must recover Eq.\,\eqref{eq: long_wavelength_omega_pl}.

\section{Fitting Details \label{app: fitting_details}}

For finding an analytical function fitting the numerical data over a broad range in the two--dimensional (\(q, r_{\scriptscriptstyle\mathrm S})-\)plane, the respective ends of the \(\omega_\mathrm{pl}(q)-\)curves deserve special care.
We therefore start with investigating the plasmon energies there.  They exhibit the following \(q-\)dependence
\begin{equation}
 \omega_\mathrm{pl}(q\!\to\!0) \>=\> a_{1/2}\, q^{1/2} + a_{3/2}\, q^{3/2} + a_{5/2}\, q^{5/2}
 \;. \label{eq: pldsp_near_0}
\end{equation}
(with appropriate coefficients \(a_{i/2}\)).  The first term is known analytically from the RPA (supplementary material, \eqref{S-Seq: pldisp qto0}); the second one, determined by the compressibility Eq.\,\eqref{eq: long_wavelength_omega_pl} is related to the correlation energy.  
Compared to the GRPA results, including 2-pair fluctuations does not modify it perceptibly. 
The coefficient \(a_{5/2}\) is obtained numerically and irrelevant here, as later replaced by the parameters given below.

In the vicinity of \(q_\mathrm{c}\) the plasmon dispersion can be modelled with a polynomial of third degree
\begin{equation}
 \omega_\mathrm{pl}(q\!\to\!q_\mathrm{c}) \>\approx\> \omega_\mathrm{pl}(q_\mathrm{c})
 + a_{\mathrm{c1}}\,q + a_{\mathrm{c2}}\,q^2 + a_{\mathrm{c3}}\,q^3 
 \;. \label{eq: pldsp_near_qc}
\end{equation}
Again, the prefactors are found numerically and then used to determine the best parameters of the overall fit.

For the order of the Pad\'e type approximation \eqref{eq: plasmon_fit pade} we tested several model complexities \(n+m\); the restriction \(n\!+\!m=\!6\) turned out as sufficient. 
The combination \(\lbrace n,m \rbrace=\lbrace 6,0\rbrace\) works best for the widest density regime, \(r_\mathrm{s} \in \lbrace 3, 30\rbrace\). 
For an ultra-high-density fit (\(r_\mathrm{s} \!\gtrsim\!30\)), shown in the supplementary material, \(\lbrace n,m \rbrace=\lbrace 2,4\rbrace\) is used. 

Based on a power expansion of Eq.\,\eqref{eq: plasmon_fit pade} in \(q\) and demanding that the just discussed limits are obeyed then yields the parameters \(p_i\,,\tilde{p}_j\).  Although solving the 6 equations for these 6 unknowns is possible with computer-algebra programs, the result is 
quite cumbersome. 
In order to achieve a more practical result, these coefficients were fitted in a second step to the density parameter \(r_\mathrm{s}\) via the ansatz:
 \begin{equation}
  p_i =\> c_0^{(i)} + c_1^{(i)}\,r_{\scriptscriptstyle\mathrm S} + c_{3/2}^{(i)}\,r_{\scriptscriptstyle\mathrm S}^{3/2} +
          c_2^{(i)}\,r_{\scriptscriptstyle\mathrm S}^2 
 \label{eq: pij_rs}
 \end{equation}
Interested readers can find details on the procedure in the Supplementary Material. 
The results for the \(c_j^{(i)}\) for (\(\lbrace n,m \rbrace=\lbrace 6,0\rbrace\)) are listed in table \ref{tab: Meta}.
The excellent agreement of the fit and the numerical data is seen in Fig.\,\ref{fig: fit_data_comp}.

\begin{table}[h]
\scriptsize
\centering
\begin{tabular}{cllll}\toprule
\(p_i\)   & \hspace{20pt}\(c_0^{(i)}\) & \hspace{20pt}\(c_1^{(i)}\)\ & \hspace{20pt}\(c_{3/2}^{(i)}\)\ & \hspace{20pt}\(c_2^{(i)}\)\  \\ \midrule
\(p_1\)   & +  0.177093     & \(-\) 0.0853141 & +  0.0159373  &  \(-\) 0.00115804\\
\(p_2\)   & \(-\) 0.159385     & + 0.0353782 & \(-\) 0.00217463 &  +  0.00371048\\
\({p}_3\) & +  1.35441     & + 0.14042   & \(-\) 0.148764   &  +  0.0252939\\
\({p}_4\) & +  2.13343     &\(-\) 0.782333  & +  0.414165   &  \(-\) 0.0501868 \\
\({p}_5\) & +  1.26389     & +  0.847607  & \(-\) 0.368082   &  +  0.0396325\\
\({p}_6\) & \(-\) 0.0800776   & \(-\) 0.273953  &  + 0.10746    &  \(-\) 0.0109972 \\ \bottomrule
\end{tabular}
 \caption{Meta--Parameters for the plasmon fit given in Eq.\,\eqref{eq: pij_rs}.
 \label{tab: Meta}}
\end{table}

\section{Plasmon dispersion coefficients}
\label{app: Plasmon dispersion coefficients}

Conceptionally, \textit{non-local} means quantities at space point \(\bf r\) depend on changes of the electromagnetic fields at \(\bf r'\). 
Mathematically a convolution in homogeneous systems and in Fourier space manifest as \(q-\)dependent response functions, this results in a dispersive \(\omega_\mathrm{pl}(q)\) \cite{WaPK11:foundations}.
While \(\omega_0^{\mathrm{\scriptscriptstyle 3D}}\!=\,\)const, \(\omega_0^{\mathrm{\scriptscriptstyle 2D}}\!\propto\!\sqrt{q}\,\) is intrinsically non-local.
Certainly, all higher order contributions to \(\omega_{\mathrm{pl}}(q)\) are dispersive, too.

By definition, \textit{correlations} are all effects beyond independent particle properties (an RPA calculation yields a DFT  exchange--correlation energy).
Calling \(\xi^{\scriptscriptstyle\mathrm{RPA}}\) `the non-local part' and terms beyond RPA `the correlation part'
is customary, but historically motivated only.

\section{Bare RPA finite width dispersion}
\label{app: Bare RPA finite width dispersion}

In realistic quantum wells\footnote{%
Here, as in \cite{hirjibehedin2002evidence,EPDH00:collective}, the same background \(\varepsilon_{\scriptscriptstyle\mathrm{b}}\!=\!13\) is assumed in the well and its surroundings.}
the confining potential is determined self-consistently with the lowest subband (,,envelope'') wave function \(\varphi_{_{0\!}}(z)\), which then may depend on \(r_{\scriptscriptstyle\mathrm S}\). 
If this effect is weak, \(\varphi_{_{0\!}}(z)\) in Eq.\,\eqref{eq: vwidth} for
\(F(q)\) can be modelled by a density-independent analytical function.
The 2D Coulomb potential \(v(q)/\varepsilon_{\scriptscriptstyle\mathrm F}\) in \eqref{eq: wpl_GRPA}--\eqref{eq: wpl_GRPA qto0} does not depend on \(r_{\scriptscriptstyle\mathrm S}\) either, so that the bare RPA finite \(L\) dispersion reads
\begin{align}
  \frac{\omega^{L,0\!}_\mathrm{pl}(q)}{\omega_{0\!}(q)} &=\> 
   \Big(1+\frac{qa_{\scriptscriptstyle\mathrm B}^*}{2F(q)} \Big)
   \Big(\frac{F(q)}{1+ \frac{qa_{\scriptscriptstyle\mathrm B}^*}{4F(q)}} \,+\,
        \frac{(qa_{\scriptscriptstyle\mathrm B}^*)^3\,r_{\scriptscriptstyle\mathrm S}^2}{8} \Big)^{\!1/2}
   \nonumber\\
   a_{\scriptscriptstyle\mathrm B}^* &=\frac{m_0\varepsilon_b}{m}\,a_0,
   \;\;a_0\!=5.29\,10^{-11}\,\mathrm{m}
   \;.  \label{eq: wpl_LRPA}
\end{align}
Clearly, the density dependence on the rhs is \({\cal O}(q^3)\) only. 
Using the spatial variance \(\Delta z^2\) of \(\varphi_{_{0\!}}(z)\), any form factor obeys
\begin{subequations}
\begin{align}
   F(q\!\to\!0) &=\>  1 - q\,\xi^{L,0}_{\scriptscriptstyle\mathrm{width}} 
                     + (q\Delta z)^2 
   \;, \label{eq: Fqto0}
 \\
  \xi^{L,0}_{} &=\>
   \int\!\!dz_1\! \int\!\!dz_2\> \big|\varphi_{_{0\!}}(z_1)\big|^2 \,
     |z_{12}|\,\big|\varphi_{_{0\!}}(z_2)\big|^2
   \;. \label{eq: xi L}
\end{align}
\end{subequations}
If for a given sample the leading RPA (`non-local') and finite \(L\) dispersion correction cancel, \(\xi^{\scriptscriptstyle L,0}\!= 3a_{\scriptscriptstyle\mathrm B}^*/8\), then
\begin{equation}
  \frac{\omega^{L,0}_\mathrm{pl}(q)}{\omega_{0\!}(q)} \approx\> 
   1+q^2\,\Big(\frac{{a_{\scriptscriptstyle\mathrm B}^*\!}^2}{32} + \frac{\Delta z^2}2 \Big)
    + {\cal O}(q^3,r_{\scriptscriptstyle\mathrm S}^2)
   \;. \label{eq: wpl_LRPA qto0}
\end{equation}
For any confining potential, we define \(L_0\) as this particular width (\(L_0\!= 9a_{\scriptscriptstyle\mathrm B}^*/(4\!-\!15/\pi^2)= 3.6a_{\scriptscriptstyle\mathrm B}^*\) in an infinite square well; \(37\!\cdot\!10^{-7}\)cm for the samples under consideration.)

\begin{figure}[h]
 \centering
 \includegraphics[width=0.45\textwidth]{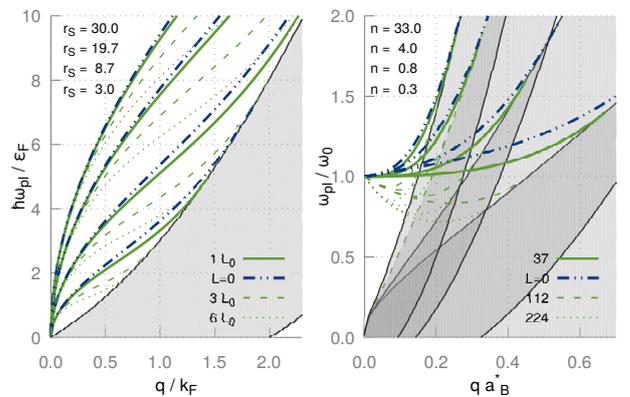}
 \caption{\label{fig: finLbareRPA}
  RPA plasmon with \(\varphi_{_{0\!}}(z)\) of an infinite square well of width \(L\!=\!0\) (dashed-dotted blue lines), \(L_0,\, 3L_0\) and \(6L_0\) (green full, dashed and dotted lines, respectively), corresponding to 370\AA,\, 1120\AA\ and 2240\AA\ in AlGaAs.
  Left: \(\omega_\mathrm{pl}(q)\) in Fermi units; the highest (lowest) group of curves are the highest (lowest) \(r_{\scriptscriptstyle\mathrm S}\).
  Right: Ratio with the classical \(\omega_0(q)\!\propto\!\sqrt{q}\,\); highest (lowest) group of curves: highest (lowest) density \(n\) (in \(10^9\)cm\(^{-2}\)).
  Shaded: particle--hole band(s).
}
\end{figure}

Figure \ref{fig: finLbareRPA} compares \(\omega^{L,0\!}_\mathrm{pl}(q)\) for different \(L\) and four of the densities shown in Fig.\,\ref{fig: fit_data_comp}\,.
While indeed hardly distinguishable from the classical dispersion over a wide \((q,r_{\scriptscriptstyle\mathrm S})-\)range for \(L\!=\!L_0\), the RPA (``non-local'') correlations are clearly overcompensated for larger \(L\).
The deviation from \(\omega_0(q)\) increases with \(q\), also for \(L_0\).

\begin{figure}[h]
 \centering
 \includegraphics[width=0.5\textwidth]{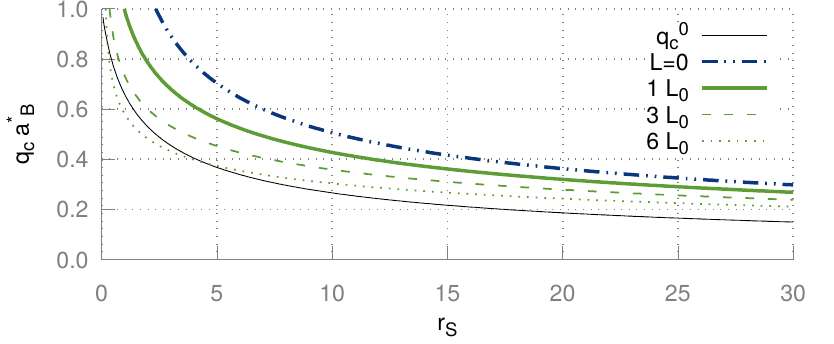}
 \caption{
  Critical wave vector for Landau damping for the QWs of Fig.\,\protect{\ref{fig: finLbareRPA}} (same line styles).
  The thin black line is where \(\omega_0(q)\) hits the particle--hole band.  \label{fig: finLbareRPAqc}
}
\end{figure}

To get a quantitative estimate how large \(q\) are influenced by the well width, Fig.\,\ref{fig: finLbareRPAqc} shows the critical \(q_c^{L,0}\) for \(L\!=\!0,\, L_0,\, 3L_0\) and \(6L_0\). 
Note that, although \(\omega^{L_0,0\!}_\mathrm{pl}(q)\) is nearly identical with \(\omega_0(q)\) for small and intermediate \(q\), the difference in \(q_c\) is substantial. 
This is a consequence of the tangential approach of the (G)RPA plasmon to the single--particle band. 
It demonstrates that for larger \(q\) correlations are more important and differently affected by \(L\) than for small ones.

We conclude with an expression for the \(L\!=\!0\) RPA critical wave vector,
\begin{equation}
 q^{0,0}_\mathrm{c}\, a_{\scriptscriptstyle\mathrm B}^* \>\approx\> 
 \frac{2\,\big(2^{1/6} \!- r_{\scriptscriptstyle\mathrm S}^{1/3} + 2^{9/4}\,r_{\scriptscriptstyle\mathrm S}\big)}
      {r_{\scriptscriptstyle\mathrm S}^{1/3}\,\big(1+ 2^{5/2}\,r_{\scriptscriptstyle\mathrm S}\big)}
 \;,
\end{equation}
which has an error of \(\le\!2\%\) except for very small \(r_{\scriptscriptstyle\mathrm S}\).

\end{appendix}

%
%

\bibliography{references}
\end{document}